\documentclass[aps,prd,reprint,superscriptaddress,nofootinbib, floatfix]{revtex4-2}

\usepackage{graphicx}
\graphicspath{ {./Figures/} }
\usepackage{amsmath,amssymb}
\usepackage{hyperref}
\usepackage{color,soul}
\usepackage{makecell}
\usepackage{orcidlink}

\begin{document}

\title{Machine Learning to assess astrophysical origin of gravitational waves triggers.}

\author{Lorenzo \surname{Mobilia}\, \orcidlink{0009-0000-3022-2358}}
\affiliation{Università degli Studi di Urbino Carlo Bo}
\affiliation{INFN, Sezione di Firenze}

\author{Gianluca Maria \surname{Guidi}\, \orcidlink{0000-0002-3061-9870}}
\affiliation{Università degli Studi di Urbino Carlo Bo}
\affiliation{INFN, Sezione di Firenze}

\begin{abstract}
In this work, we explore a possible application of a machine learning classifier for candidate events in a template-based search for gravitational-wave (GW) signals from various compact system sources.	
	 We analyze data from the O3a and O3b data acquisition campaign, during which the sensitivity of ground-based detectors is limited by real non-Gaussian noise transient. The state-of-the-art searches for such signals tipically rely on the signal-to-noise ratio (SNR) and a chi-square test to assess the consistency of the signal with an inspiral template. In addition, a combination of these and other statistical properties are used to build a 're-weighted SNR' statistics. We evaluate a Random Forest classifiers on a set of double-coincidence events identified using the MBTA pipeline. The new classifier achieves a modest but consistent increase in event detection at low false positive rates relative to the standard search. Using the output statistics from the Random Forest classifier, we compute the probability of astrophysical origin for each event, denoted as $p_\mathrm{astro}$. This is then evaluated for the events listed in existing catalogs, with results consistent with those from the standard search. Finally, we search for new possible candidates using this new statistics, with $p_\mathrm{astro}  > 0.5$, obtaining a new subthreshold candidate (IFAR =0.05) event at $gps: 1240423628$ .
\end{abstract}

\maketitle

\section{Introduction}
Following the detection of the first gravitational waves signal GW150914 \cite{gw150914}, by the LIGO-Virgo collaboration, gravitational-wave astronomy has become a reality. Since then, more than 200 events have been observed \cite{gwtc1, gwtc2, gwtc3, gwtc4}. The discovery of the first binary neutron star system, GW170817  \cite{gw170817},  marked the advent of multi-messenger astrophysics, an area of significant interest for both cosmology \cite{Hubble} and fundamental physics \cite{tidalDeform}. 
	These scientific breakthroughs have been made possible through the operation of Advanced LIGO \cite{advLIGO} interferometers in the United States of America and Advanced Virgo \cite{Virgo}, in Italy. Looking ahead, the next generation of interferometers -  such as Einstein Telescope \cite{ET} in Europe, Cosmic Explorer \cite{CE} in the USA, and space-based detectors like LISA \cite{LISA} - will offer unprecedent insight into gravitational physics and further enhance our understanding of the universe.
	
	The data stream provided by the interferometers during data acquisition campaigns is analysed using different pipelines, depending on the target source class. These include pipelines for compact binary coalescence (CBC) searches \cite{pycbc1, gstlal1, mbta1} and burst analyses \cite{PhysRevD.100.042003, cwb, PhysRevD.107.062002}. The formers focus on identifying astrophysical signals from sources such as binary black holes (BBHs), neutron star-black hole (NSBH) systems and binary neutron stars (BNSs), where the waveform is known \textit{a priori}. In this case the matched-filtering technique \cite{matched_filtering} can be applied, enabling an optimal search strategy under the assumption of Gaussian noise.
	 
	 In practice, however,  the presence of transient noise artifacts --- known as glitches --- can lead to false triggers of non-astrophysical origin. Developing robust methods to improve the discrimination between noise and genuine signals is therefore essential to ensure the statistical reliability of candidate gravitational-wave events. 
	 Supervised machine learning offers a promising approach \cite{MLpycbc, MLgstlal}: classifiers can be trained to exploit features extracted from the pipeline triggers to distinguish between noise artifacts and true gravitational-waves signals. Several attempts using machine learning methods have been made in this context, as the aforementioned works describe, aiming to enhance the separability of the two populations and improve overall performance. Furthermore, supervised algorithms have been applied to re-weight the ranking-statistics, and improve the signal-noise classification also in burst searches. For example, \cite{cWB-ml1} introduces a decision tree learning algorithm XGBoost and \cite{GMM} a Gaussian mixture model to classify Coherent WaveBurst triggers, and both methods successfully increase the search sensitivity towards short-duration GW in O3 data \cite{cWB-ml2, cWB-ml3}
	
	This article builds upon the research direction established by the aforementioned works  \cite{MLpycbc, MLgstlal}. These previous studies have primarly focused on the analysis of single-detection triggers, i.e. triggers provided by compact-binary-coalescence pipelines originating from individual interferometers. Such investigations demonstrated a clear improvement in detection performance of gravitational-wave pipelines through the application of machine learning techniques.
	In this work, we shift the focus from single-detector triggers to coincidence triggers, specifically those in temporal coincidence between the Hanford (H) and Livingston (L) interferometers, as identified by the Multi-Band Template Analysis (MBTA) pipeline \cite{mbta2, mbta3} during the O3 observing run \cite{O3data}. A supervised machine learning classifier is trained using a set features extracted from these coincident triggers.
	The resulting detection statistic from the classifier is compared with the standard ranking statistic used by the pipeline, and shows a compatible result in detection sensitivity. As a further application, we compute the probability that a trigger is astrophysical --- denoted as  $p_\mathrm{astro}$ \cite{pAstro1} --- using the new statistic. We compare this result with respect to the classical  $p_\mathrm{astro}$ computation based on the original ranking statistics provided by the MBTA pipeline. \cite{pAstroMBTA}. We want to highlight that the choice of MBTA was done for convenience, due the authors currently work in the pipeline. In addiction, this study has not been deployed for MBTA pipeline specifically, but, due to its generality, can be applied easily to other pipelines that rely their search on matched-filtering analysis. 
	
	The article is structured as follow: in Sec. II, the MBTA pipeline is briefly described, highlighting its fundamental working principles; in Sec. III, the supervised machine learning algorithm, Random Forest, is introduced, and the detection statistic derived from it is defined; in Sec. IV, the features used and the hyperparameters selected for the algorithm are presented, along with the procedure adopted for their selection; in Sec. V, the results and detection capabilities of the Random Forest classifier are compared with those obtained using the stardard MBTA statistic;  in Sec. VI, a possible application of the new statistic is explored by integrating it into the definition of $p_\mathrm{astro}$; in Sec VII, the $p_\mathrm{astro}$ values derived from the Random Forest-based statistic are compared with those obtained using the MBTA ranking statistics, focusing on events reported in the GWTC-2 \cite{gwtc2} and GWTC-3.0 \cite{gwtc3} catalogs; in Sec. VIII, a full search across the O3 dataset is performed using this new statistic to identify potential new candidate events; finally, conclusions and future perspectives are discussed in Sec. IX. 
	In appendix A the features distributions for noise and injection population are depicted. 

\section{The MBTA Pipeline}
The Multi-Band Template Analysis pipeline is an algorithm developed for the compact binary coalescence (CBC) searches and it is currently in use during the O4 data acquisition campaign. The O3 version of the MBTA pipeline, used in this work, is fully described in \cite{mbta2}. Here, we highlight only the key features relevant to the present study. This pipeline is based on the matched-filtering technique, which is applied to the data collected by the interferometers. To employ matched filtering, a set of model gravitational wave signals --- known as templates --- is required. These templates are organized into a template bank that spans the parameter space of potential astrophysical sources. 
	A distinctive feature of MBTA is its use of multiple frequency bands:  the matched filtering is computed in parallel over separate frequency ranges. This multi-band approach enables a consistent reduction in the number of templates needed, thereby improving computational efficiency, an essential aspect for real-time, online analysis.
	 Since a substantial fraction of the pipeline's triggers originates from transient-noise artifacts (glitches), synthetic gravitational-wave signals, known as injections, are added directly to the data stream. This allows the construction of a statistical distribution for both glitch-induced and injection-induced triggers. If a real candidate exhibits statistical properties similar to those of the injection population, it can be considered a potential astrophysical event.
	 The output of the match-filtering  process is the signal to noise ratio (SNR), denoted as $\rho$. Matched filtering is first applied independently to each interferometers data stream. If the SNR exceeds a predefined threshold $\rho_{\text{min}}$, a trigger is recorded. 
	 To enhance the separation between noise and signal population, a ranking statistic is build based on the SNR distribution. This statistic penalizes $\rho$ in the presence of features tipically associated with glitches. It is computed using an autocorrelation-based least-squares test $\xi^2_{PQ}$, as described in \cite{messick_autochi}, which quantifies the similarity between the observed $\rho$ time series and the expected from a genuine signal. The resulting re-weighted SNR, denoted as $\rho_{\text{rw}}$, accounts for this deviation.
	 In addition, in the O3 MBTA version, to monitor periods when a detector exhibits unusually high glitch activity, an observable called Excess Rate (ER) is introduced. This measures the excess in rate, around the trigger, of initial triggers relative to the rate of triggers that survive after applying a cut on $\rho_{\text{rw}}$. The combination of  $\rho_{\text{rw}}$ and ER defines a new statistic $\rho_{\text{rw, ER}}$, which further improves the discrimination between signals and noise by down-ranking triggers associated to noisy periods. This observable has been discarded in the O4 pipeline version, see \cite{mbta3}. 
	 Once $\rho_{\text{rw, ER}}$ is computed for individual triggers, coincidences triggers are identified by comparing their GPS times. 
	 The pipeline then outputs several parameters associated with this trigger, including $\rho$ as well the masses and spins of the corresponding template. A pair of triggers from Hanford (H) and Livingston (L) are considered coincidence (an HL event) if their time difference is within $15ms$, accounting for the light-travel time between the detectors. True astrophysical signals are expected to exhibit correlations across detectors, not only in their arrival times but also in phase and amplitude.
	 The final ranking statistic for coincident events, known as amplitude or $\rho_{RS_{HL}}$, incorporates this multi-detector consistency. It is computed by combining amplitude-related information described in \cite{Nitz}, applying corrections to reflect the expected coherence of genuine signals. Finally, a cut in $\rho_{RS_{HL}} $ is applied to retain candidate events.

\section{Random Forest Algorithm and Training Procedure}
	
\begin{figure}
	\includegraphics[width=0.5\textwidth]{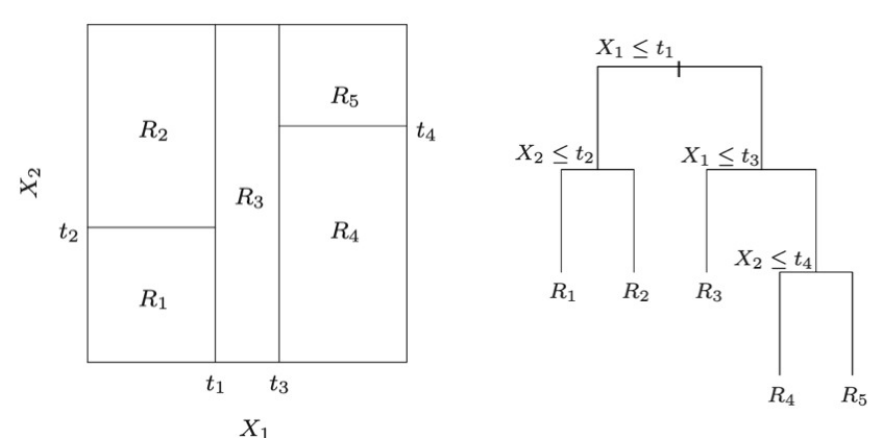}
		\caption{Tree-based decision algorithm. The features $X_1$ and $X_2$ define a two-dimensional feature space. On the left, the threshold $t_i$ represents the decision rules used to partition this space into rectangular regions $R_i$. On the right, the corresponding decision tree representation is shown, where each internal node represents a binary decision based on a threshold, and each terminal node (leaf) correspond to a region $R_i$ in the feature space. Credits to \cite{statisticalLearningBook}.}
		\label{fig:RF}
\end{figure}

	The Random Forest algorithm \cite{breiman1} is a widely used and flexible supervised learning method, primarily employed for classification and regression tasks. It classifies a data point by estimating the probability of belonging to a specific class, based on the analysis of  multiple independent --- partially correlated --- input variables known as 'features'.
	At the core of the method there is a decision tree structure \cite{statisticalLearningBook}. Each tree is constructed through a series of binary decisions: at each node, a feature and a corresponding threshold (known as decision rule) are selected to maximize the separation between classes. The tree continues to split recursively until it reaches leaves, which represent the final class predictions or regression outputs for subsets of the data. The scheme of this procedure is reported in Fig. \ref{fig:RF}.
	A Random Forest consists of an ensemble of such trees, each trained independently. To introduce variability and reduce overfitting, the algorithm employs the bootstrap aggregation (bagging) technique: each tree is trained on a randomly resampled subset of the original dataset, and at each split, a random subset of features is considered for the decision rule. This controlled randomness helps mitigate bias and variance, improving generalization performance.  After training on a labeled dataset, the performance of the Random Forest is evaluated on unlabeled test data.

In our work, we focus on a classification problem using a probabilistic classifier. For each data in the training dataset, the forest processes the input through all trees. In each tree, this example undergoes a decision process and reaches a specific leaf node. During training, each leaf records the proportion of training samples from each class that fell into it. This proportion is used as the per-tree probability score for the input datum.
	The final score assigned by the forest is the average of per-trees scores, resulting in a value in the range $[0,1]$, which represents the estimated probability that the event belongs to a Signal class (as opposed to Noise).
	More formally, given a forest composed of $N$ trees, where each tree $i$ outputs a score $p_{s,i} \in [0,1]$ for a given data point, the final classification score is computed as 

\begin{equation}
	p_{s} = \frac{1}{N}\sum_{i=1}^{N} p_{s,i}
\end{equation}

	To be effective, machine learning algorithms of this kind require a large volume of data. However, since only a few tens of gravitational-wave events are expected during a typical observing run,  we cannot rely solely on real astrophysical signals. Therefore, synthetic injections are used to assess the pipeline sensitivities. These injections consist of software-generated waveforms, corresponding to model-predicted gravitational-wave signals, which are directly 'injected' into the detector data stream.
	When the MBTA pipeline analyzes the data, it possibly produces trigger at the injection time, which is labeled as signal. This procedure enables the construction of a statistically robust set of signal-class triggers. By injecting a sufficiently large number of simulated events, we can train the Random Forest to distinguish noise-induced artifacts (glitches) from genuine signal-like events.  
	The expectation is that, once trained, the classifier will recognize the characteristic features of real signals and assign them high signal-like scores, while down-ranking glitchy events. The injections used in this study are those performed during the O3 observing campaign, and are drawn from a population model designed to approximate the expected distribution of true astrophysical sources, as described in \cite{gwtc2, gwtc3}. In particular, the BBH injections have been produced considering a truncated power law for $m_1$, spanning $m_{1, \mathrm{min}} = 2M_\odot$ and $m_{1, \mathrm{max}} = 100M_\odot$ considering $m_2 < m_1$. The injections were generated using the approximant\texttt{SEOBNRv4PHMpseudoFourPN} \cite{PhysRevD.102.044055}. Also, spins have been considered isotropically distributed with max value of $0.998$. For BNS systems the masses are uniformly distributed within range $m \in [1, 2.5]M_\odot$. Similarly, the spins follow the same distribution rule with a maximum allowed spin of $0.4$. The approximant used is \texttt{IMRPhenomPv2} \cite{PhysRevD.93.044006}. The NSBH systems are obtained considering uniform distribution for masses and spins. The component masses are in ranges $m_1 \in [1, 2.5]M_\odot$ and  $m_2 \in [2.5, 60]M_\odot$, the spins have maximum values of $0.4$ and $0.998$ respectively. The approximant considered is \texttt{IMRPhenomPv2} \cite{PhysRevD.93.044006}

\section{Dataset, Features Selection, and Hyperparameters Optimization}

	In this work, the dataset consists of triggers produced by the MBTA pipeline during the O3 data acquisition campaign, which is divided into two sub-runs: O3a (1 April 2019 15:00 UTC  - 1 October 2019 15:00 UTC) and O3b (1 November 2019 15:00 UTC - 27 March 2020 17:00 UTC). 
	We focus on HL-Von coincidence triggers, defined as those produced by MBTA when all the three interferometers --- Hanford (H), Livingstone (L) and Virgo (V) --- were simultaneously operational (i.e. in science mode and acquiring data). Coincidences are identified when both H and L triggers satisfy the condition $\rho_{H}, \rho_{L} > \rho_{\text{min}} = 4.5$, and the time difference between the two falls within the predefined coincidence window. For these events,  the SNR of Virgo $\rho_V$, is also recorded.
	Triggers are labeled as injection-associated if a pipeline trigger occurs within the window $[-80, +40]$ms of the injection time. Among all such triggers, the loudest one (i.e., the one with the highest $\rho$) is selected and labeled as a signal trigger.
For real astrophysical events, we use the double-coincidence HL triggers provided by MBTA and reported in the GWTC-2.1 and GWTC-3.0 catalogs. The corresponding trigger counts and dataset composition are summarized in Table I.
	
\begin{table}[t]
\caption{HL–VOn coincidence triggers from the MBTA pipeline during the O3 observing run, restricted to times when Hanford, Livingston, and Virgo were simultaneously operational.}
\label{tab:dataset}
\centering
\begin{ruledtabular}
\begin{tabular}{lccc}
 & Noise & Injections & Astrophysical events \\
\colrule
O3a & 173\,130 & 83\,258 & 23 \\
O3b & 129\,133 & 76\,389 & 16 \\
\end{tabular}
\end{ruledtabular}
\end{table}

	To achieve optimal performance from a classifier, it is common practice to use balanced dataset. A balanced dataset contains an equal --- or at least compatible --- number of samples for each class. In our case, the two calsses are 'noise' and 'injections'. While it is possible to train a classifier on an imbalanced dataset, with number of injections reflecting the expected number of real events during an expecting run, such an approach tipically requires more advanced handling strategies to mitigate class imbalance effects \cite{unbalanced1, unbalanced2}. 
	To simplify the implementation and fully exploit the informations available in the dataset, we opted for a balanced dataset, using all the available injections. The data were split into training ($70\%$) and Test  ($30\%$) subset. For Random Forest algorithm the validation dataset is not required since the bootstrap sampling ensure an un-biased classificator.For O3a, the training set includes 58,281 noise triggers and an equal number of injection triggers. The test dataset consists of 24,977 noise and 24,977 injection triggers. For O3b, the training set includes 53,472 noise triggers and an equal number of injections, while the test set consists of 22,917 triggers for each class. 
	The split was performed such that there is no overlap between training and test dataset.   Additionally, data were sampled randomly from the full dataset to avoid any bias that might arise from specific detector conditions or non-stationary noise behavior at certain times. This results in two statistically independent datasets, each with a balanced representation of noise and injection triggers. The next step is to define the set of features that will be used to train the algorithm.

	As described in Sec. II, the triggers produced by the MBTA pipeline are characterized by a variety of features. Some of these can be classified as a statistical features, such as the signal-to-noise ratio ($\rho$) or the autocorrelation-based least-squares statistic ($\xi^2_{PQ}$), while others are physical features, including the component masses and spins of the template associated with the trigger.
	A Random Forest classifier trained on a dataset incorporating this diverse set of features should be capable of constructing a model that effectively distinguishes noise from signal, potentially outperforming the standard MBTA ranking statistic in terms of classification capability. For this purpose, we used the same features used by MBTA, considering also the physical ones. The Random Forest algorithm then was trained using the following set of features: 
	
\begin{itemize}
	\item $\rho$ (Signal-to-Noise-Ratio):  For HL-Von coincidence triggers, we consider $\rho_H$ and $\rho_L$, each required to satisfy $\rho > \rho_{\text{min}}$.
	\item $\xi^2_{PQ}$: The autocorrelation-based least-squares statistic \cite{messick_autochi}, computed for both Hanford and Livingston: $\xi^2_{PQ,H}$ and $\xi^2_{PQ,L}$.
	\item $ER$: The excess trigger rate, used to characterize periods of elevated noise, is included as $ER_H$ and $ER_L$.
	\item nEvents: The number of triggers within the cluster associated with the event, as defined in MBTA \cite{mbta2}. 
	\item  Component masses ($m_1$, $m_2$) and aligned spins ($\chi_1$, $\chi_2$) of the binary system (for masses distribution see Fig. \ref{fig:mass-population}). These are included to evaluate whether specific regions of the parameter space are more susceptible to noise contamination. The ranges of masses and spins are chosen to coincide with the full parameter-space coverage of the template bank used in the search. As a result, the training set samples the same distribution encountered at inference time, without introducing ad-hoc restrictions or tuning of the parameter space that could artificially enhance the classifier’s performance. To assess whether the intrinsic parameter, such as the total mass, could drive the glitch classification, we evaluated the fraction of background triggers assigned with high score $p_s$ as a function of total mass (noise leakage). While a mild increase in leakage is observed toward higher masses --- consistent with the increased susceptibility of short-duration templates to noise transients --- the effect is gradual and no single mass region dominates the background contamination.  The risk of overfitting associated with the inclusion of intrinsic parameters is mitigated by the use of an ensemble of decorrelated decision trees with feature subsampling at each split, which limits model variance. In addition, classifier performance and score distributions were validated on independent data not used during training, confirming stable behavior across the explored parameter space.
	\item $t_{\text{dur}}$:The duration of the waveform template. This choice is motivated by the expectation that many glitches are short-lived compared to astrophysical signals such as BNS. However, high-mass BBH also produce short-duration signals, comparable to glitches.
	\item $\Delta \phi$: The difference in phase; $\Delta t$: The difference in time; $\Delta D$: The difference in effective distance. All of those have been evaluated between the $H$ and $L$ triggers. These features capture inter-detector consistency and are included to allow the classifier to learn patterns associated with true coincident events versus coincidental noise triggers.
\end{itemize}


	The dataset's features distribution used in this work is detailed in Appendix~\ref{appendix:features}. As discussed in Sec. II, the majority of the features listed above have already been incorporated into the construction of the standard MBTA ranking statistic. In this work, we investigate whether a machine learning algorithm can leverage these same features --- augmented with additional information such as component masses, spins, template duration, and the number of triggers in a cluster (that it is referred to as nEvents)--- to construct a more effective classification statistic. 
The next step is to define the architecture of the algorithm.
	In general, a machine learning model includes a number of hyperparameters, which are configuration settings not learned during training but instead tuned externally to optimize model performance. The tuning process involves training several models with different hyperparameter combinations and comparing their performance on the training dataset. The tuning and the definition of the hyperparameters depend on the library considered.
To evaluate and compare different models, a suitable performance metric is required. In this work, we adopt the $F_1$ score as our evaluation metric. The $F_1$ score is defined as the harmonic mean of precision and recall \cite{f1-score}, and is given by:

\begin{equation}
	F_1 = \frac{2TP}{2TP + FP + FN}
\end{equation}

	Where true positives (TP) and true negatives (TN) refer to the number of correctly labeled injection and noise triggers, respectively, while false positives (FP) and false negatives (FN) correspond to incorrectly labeled injection and noise triggers. These quantities are evaluated using a classification threshold of $p_s = 0.5$. The model classifies a trigger as 'signal' if its signal-like probability exceeds 0.5, and as 'noise' otherwise.
Each model, once trained on a subset of the training dataset, is evaluated on a validation split from the same training set. The resulting $F_1$ score reflects the model’s ability to balance precision and recall, and serves as a proxy for its overall classification accuracy. In this metric, well-performing models yield scores close to 1, while poor models tend toward 0.
The algorithm performs a hyperparameter search across the configuration space. For each combination of hyperparameters, the model is trained and its $F_1$ score is evaluated. The set of hyperparameters that achieves the highest $F_1$ score is selected as optimal.
 In this setup, each decision tree in the forest acts as a binary classifier rather than assigning a continuous probabilistic ranking. Even though we chose this metrics to optimize the set of hyperparameters, this is not the solely possible solution. Several alternative metrics are available for evaluating model performance in the current state-of-the-art literature, such as accuracy, defined as $\text{Accuracy} = \frac{TP + TN}{TP + TN + FP + FN}$ \cite{accuracy} and the 'Jaccard index', $\text{J} = \frac{TP}{TP + FP + FN}$, \cite{Jaccard}. The choice of the $F_1$ score in this work is motivated by its ability to quantify the separability between classes, while incorporating similar information to that captured by accuracy. Additionally, $F_1$ is more commonly used than the Jaccard index in binary classification tasks, and is widely regarded as a standard metric when both false positives and false negatives are of concern.  The package used in this work is \texttt{scikit-learn} \cite{scikit-learn} open source package. The \texttt{typewriter} font identifies the build-in hyperparameters provided by \texttt{scikit-learn}. The hyperparameters used are

\begin{itemize}
	\item \texttt{n$\_$Estimators}: The number of trees that constitutes the forest. Increasing this number will cause a better performance since the errors are averaged away. On the other side, a huge number of trees will cause a consistent computational time increase. Also, the performances will not gain precision after a certain value of trees, so choosing the best number of tree is done in order to maximise the precision and minimize the computational time. 
	\item \texttt{criterion}: Is the rule used by the tree to split its nodes. The rules usable are \texttt{gini} or \texttt{entropy} . The former is the Gini impurity function, while the latter is the Shannon information gain.
	\item \texttt{max$\_$depth}: This hyperparameters defines how in depth a single tree can grow in order to obtain the best-purity ending nodes. It has been proved that over-grown trees tend to overfitting, so put a limit in this value is fundamental to get a properly-working classifier. On the other side, a small number of trees may not be able to learn the structures in the data.
	\item \texttt{min$\_$samples$\_$leaf}: The minimum number of samples required to be at a leaf node: a split node at any level of the tree can be considered only if this number of training samples fall in each of the left and right branches.
	\item \texttt{min$\_$samples$\_$split}: The minimum number of samples required to split an internal node.
	\item \texttt{max$\_$features}: The number of features considered to apply the best split. In this work we used only \texttt{sqrt}, that means that given a set of $N$ features, each tree will be trained using $\sqrt{N}$ randomly chosen features. 
	\item \texttt{ccp$\_$alpha}: Complexity parameter used for Minimal Cost-Complexity Pruning. The idea behind this process is to grow the tree until a minimum node size is reached, then this tree is pruned using the Cost-Complexity Pruning function. This parameter will prune the tree choosing the most-complex one with a value of $\alpha$ below the \texttt{ccp$\_$alpha} threshold, where $\alpha$ is the parameter that rule the trade-off between the size of the tree and its capability in fitting the data. For more references see \cite{statisticalLearningBook}.
\end{itemize}

To identify the optimal combination of hyperparameters, a grid search was performed. Each predefined combination of hyperparameters was used to train and validate a separate model, and the combination that achieved the highest $F_1$ score was selected as the best configuration. The hyperparameter grid used in this search is reported in Table \ref{table:hyperparams}. The $F_1$ scores obtained for the best model are 0.968 for O3a and 0.965 for O3b.

\begin{table*}
\caption{Hyperparameter space explored during grid search for the O3 dataset. The best-performing values are reported separately for O3a and O3b.}
\label{tab:hyperparams}
\centering
\begin{ruledtabular}
\begin{tabular}{lccc}
\textbf{Hyperparameter} & \textbf{Values} & \textbf{Best O3a} & \textbf{Best O3b} \\
\colrule
\texttt{n\_estimators}        & 15, 50, 100               & 100         & 100 \\
\texttt{criterion}           & \texttt{gini}, \texttt{entropy} & \texttt{entropy} & \texttt{entropy} \\
\texttt{max\_depth}          & 10, 12, 15 & 12 & 15 \\
\texttt{min\_samples\_leaf}  & 1, 5, 10                & 1           & 1 \\
\texttt{min\_samples\_split} & 2, 5, 10                & 2           & 5 \\
\texttt{ccp\_alpha}        & None, 1e$^{-5}$, 5e$^{-5}$, 1.5e$^{-4}$ & None & None \\
\end{tabular}
\end{ruledtabular}
\label{table:hyperparams}
\end{table*}


For the O3b case, we highlight that although a deeper random-forest configuration yields a marginally higher $F_1$ score, we do not adopt it in the final analysis. 
The probability of astrophysical origin is constructed from the class score densities combined with strongly asymmetric priors, reflecting the large imbalance between expected background triggers and true astrophysical signals. As a result, the inferred probability is particularly sensitive to the behavior of the background score distribution in the high-score tail.
We find that increasing the tree depth leads to excessively sharp score distributions for the background class and to unstable behavior in the extreme tail. This, in turn, amplifies background leakage into the high-probability regime and degrades the robustness of the posterior probability assigned to rare events.
We therefore impose a maximum tree depth as an explicit regularization choice, favoring a model with slightly reduced global classification performance but substantially improved stability of the background tail, which is the regime most relevant for rare-event inference.
We verified that this behavior is not reflected in global metrics such as the ROC curve or $F_1$ score, which are dominated by the bulk of the score distribution and therefore insensitive to tail instabilities.

In addition, we investigate the robustness of the trained models across datasets. Specifically, models trained on O3a data are evaluated on the O3b dataset and vice versa, allowing us to assess cross-dataset generalization and stability. The results of this analysis are discussed in Section V.

It is also interesting to compare these values with those reported in other studies, such as \cite{CaudillRF} and \cite{HodgeRF}. The number of trees is similar across all studies, except for the ringdown-only analysis in \cite{CaudillRF}, which uses a different configuration. Furthermore, other hyperparameters explored in those works—such as leaf size and the splitting criterion—are broadly consistent with those used in this study. This supports the hypothesis that, for this class of problems, where the complexity and range of input features are comparable, the Random Forest algorithm tends to converge toward a similar model architecture.

\section{Performance Evaluation on the Test Dataset}

Once the model is trained, the test dataset is used to evaluate the performance of the algorithm. For each trigger in the test set, the model assigns a score $p_s$, representing the confidence that the trigger belongs to the 'injection' class. A value of $p_s \sim 1$ indicates that the algorithm is highly confident the trigger is signal-like, while $p_s \sim 0$ indicates it is classified as noise-like.
The resulting $p_s$ distributions for both the noise and injection populations in the test dataset are shown in Fig. 2 (b). For comparison, also the ranking-statistics distribution is illustrated in Fig. 2 (a). These distributions reveal that the algorithm does not achieve a complete separation between the two populations in $p_s$ space, indicating some overlap in classification confidence.

\begin{figure}[t]
\centering
\includegraphics[width=\linewidth]{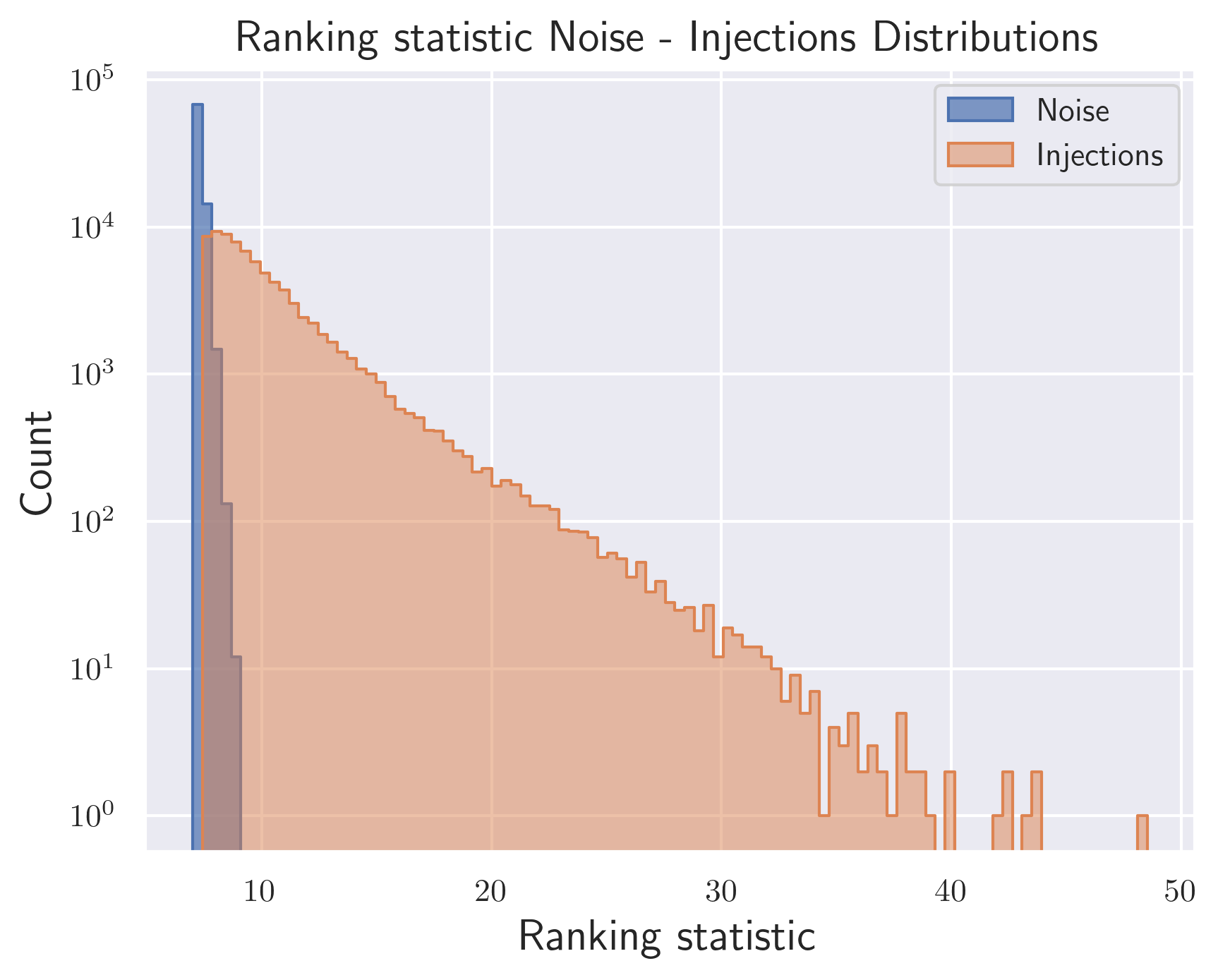}
\hfill
\includegraphics[width=\linewidth]{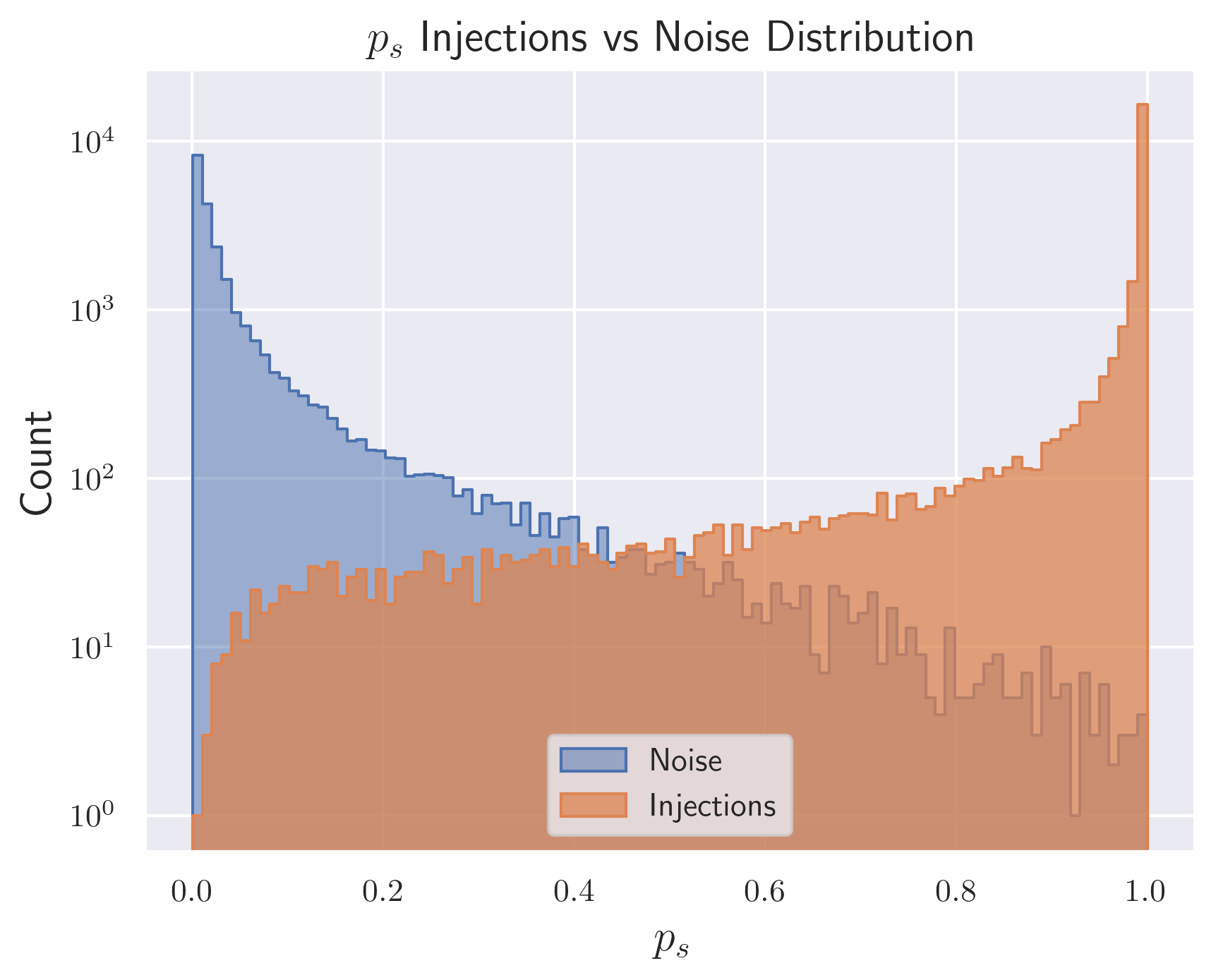}
\caption{Statistics obtained for the O3a dataset: 
(a) amplitude distribution for Noise (red) and Injections (blue), as computed by MBTA; 
(b) $p_s$ distribution for Noise (red) and Injections (blue), as produced by the classifier.}
\label{fig:figures}
\end{figure}

The False Alarm Probability $\alpha$ is defined as the fraction of noise triggers with a classifier score $p_{s,i}$, where $i$ is the data index, greater than a given threshold $\hat{p}_s$. Similarly, $N_d$ is defined as the number of injection triggers with $p_{s,i} > \hat{p}_s$. Formally, let $N_n$ denote the total number of noise triggers and $N_s$ the total number of injections. Then, the quantities are defined as:

\begin{align}
	\alpha = &  \frac{1}{N_n} \sum_{i=1}^{N_n}\theta(p_{s,i} - \hat{p}_s) \\
	N_d = & \sum_{i=1}^{N_s}\theta(p_{s,i} - \hat{p}_s) \\
\end{align}

\begin{figure}[t]
\centering
\includegraphics[width=\linewidth]{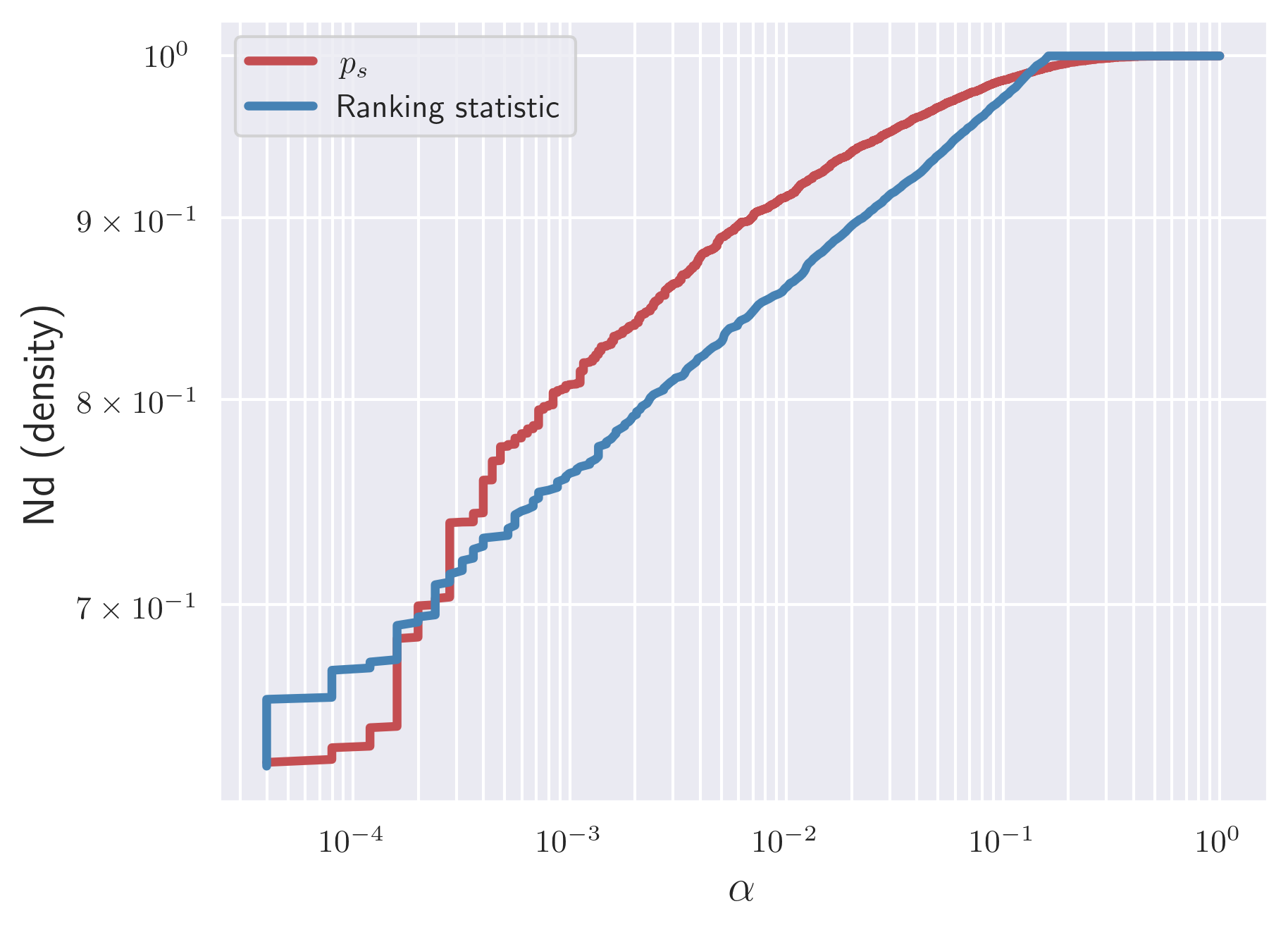}
\hfill
\includegraphics[width=\linewidth]{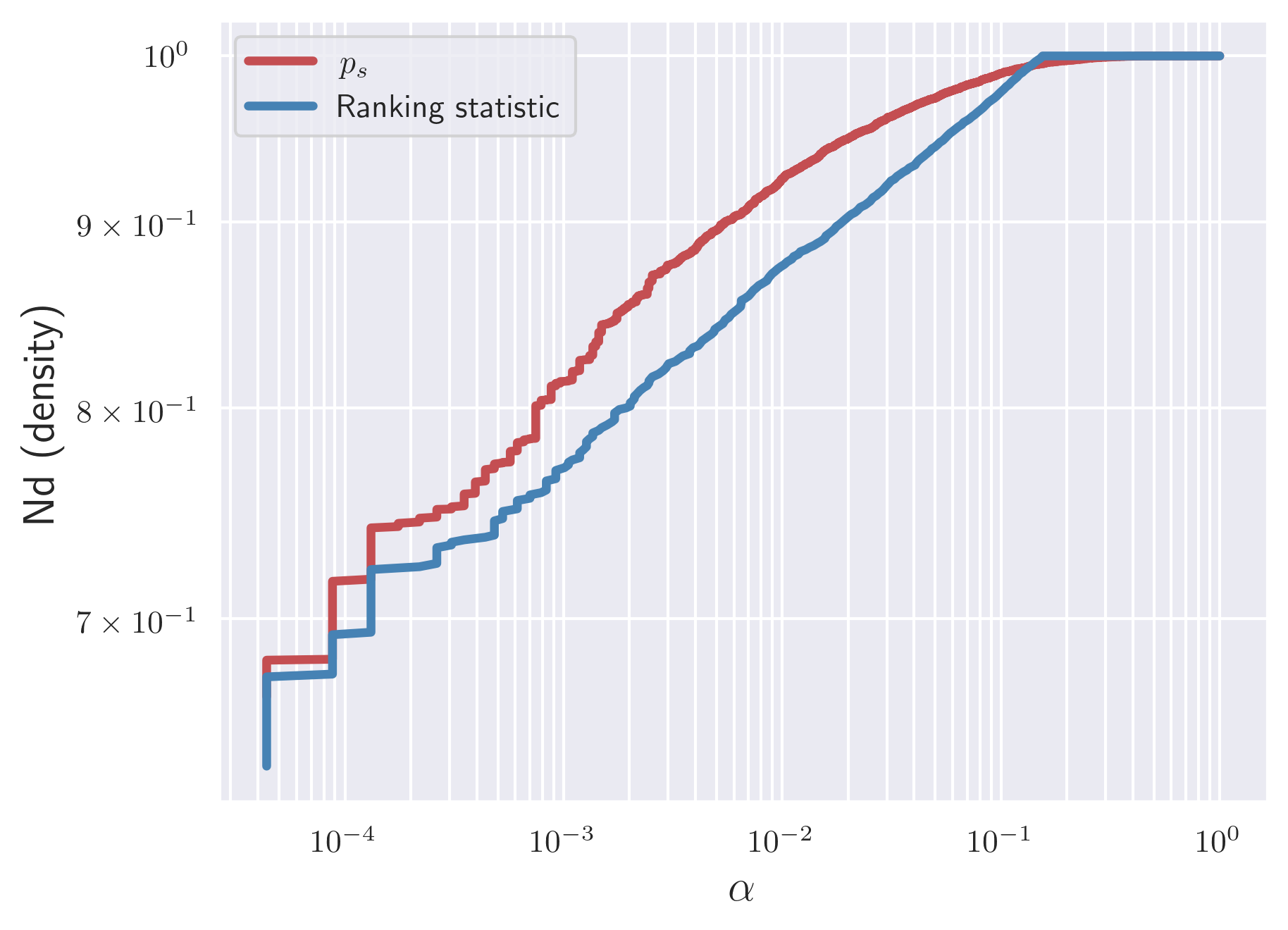}
\caption{Comparison of ROC curves between the Random Forest and MBTA pipelines for the O3 dataset:
(a) O3a observing period and (b) O3b observing period.}
\label{fig:Rocs}
\end{figure}

	Here $\theta$ denotes the Heaviside step function. If $N_d$ is expressed as function of $\alpha$, then the Receiver Operating Characteristic (ROC) curve can be constructed. This curve characterizes the trade-off between the detection efficiency and the false alarm probability as the classification threshold $\hat{p}_s$ is varied. 
	By computing the same quantities using the ranking statistic provided by the MBTA pipeline instead of the classifier score $p_s$, we can perform a direct comparison between the Random Forest classifier and the traditional pipeline-based ranking.
	The resulting ROC curve from the Random Forest is generally above the curve obtained from the MBTA ranking statistic, as reported in Fig. \ref{fig:Rocs}. This indicates that the classifier is more effective at distinguishing between noise and injection triggers, demonstrating improved classification performance over the standard pipeline approach.

\begin{figure}[t]
\centering
\includegraphics[width=\linewidth]{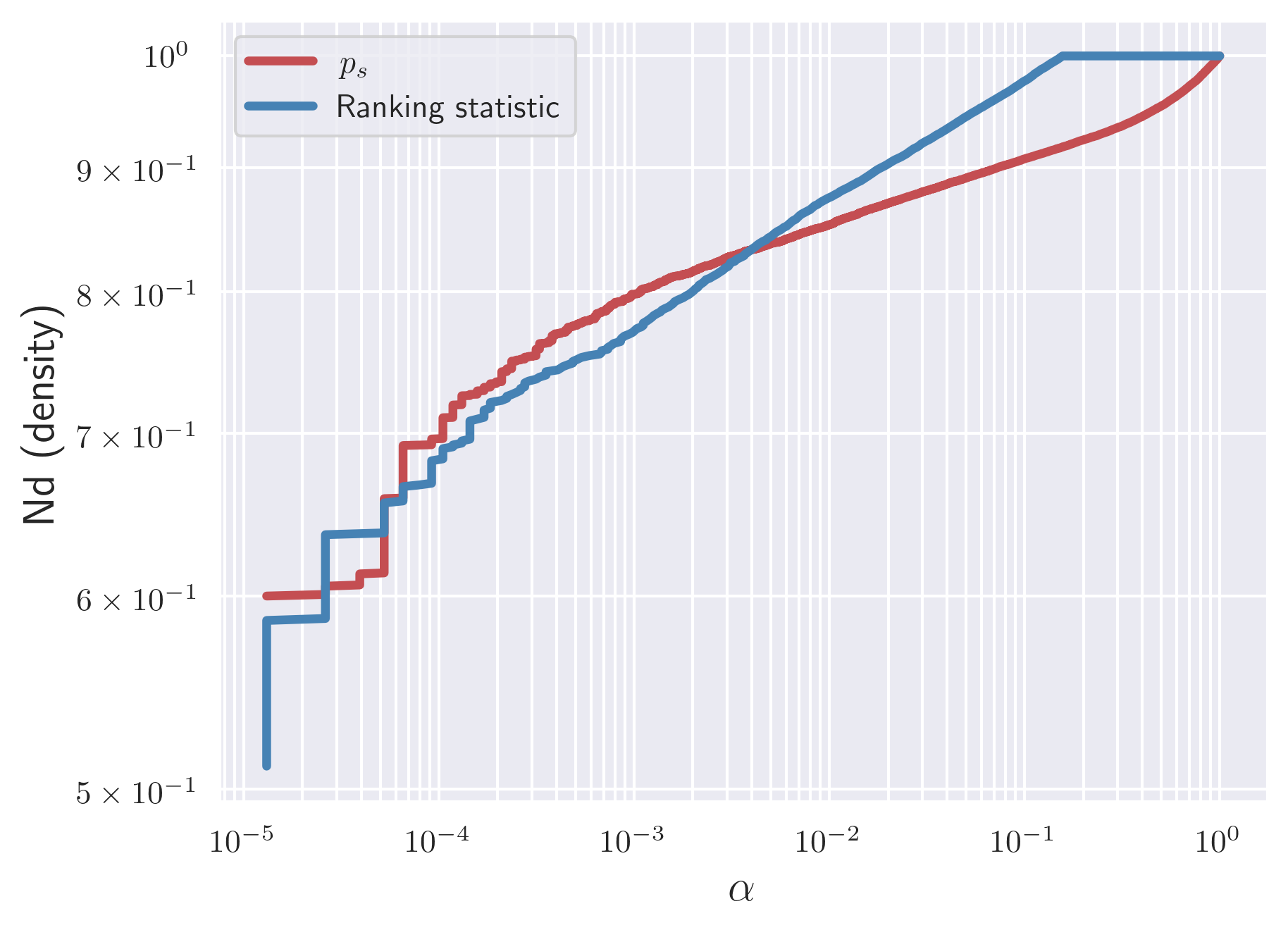}
\caption{Performance (ROC) of the Random Forest classifier trained on O3a and evaluated on O3b.}
\label{fig:roc_o3b_o3amodel}
\end{figure}

To assess the stability of the model, we applied the Random Forest trained on the O3a dataset to the O3b data. As in the previous analysis, a balanced dataset of noise and injection-labeled triggers was constructed, ensuring that all triggers associated with catalogued events were excluded.
The resulting ROC curve is shown in Fig. \ref{fig:roc_o3b_o3amodel}. In this case, the classifier achieves performance comparable to the ranking statistic produced by the MBTA pipeline. However, this also indicates that the classifier's performance deteriorates when applied to data outside its training domain, as it no longer compatible with the MBTA statistic as it did in the matched training-testing scenario. This suggests that the classifier exhibits some degree of overfitting to the specific characteristics of the O3a data and may require additional regularization or retraining for improved generalization.

\begin{figure}[t]
\centering
\includegraphics[width=\linewidth]{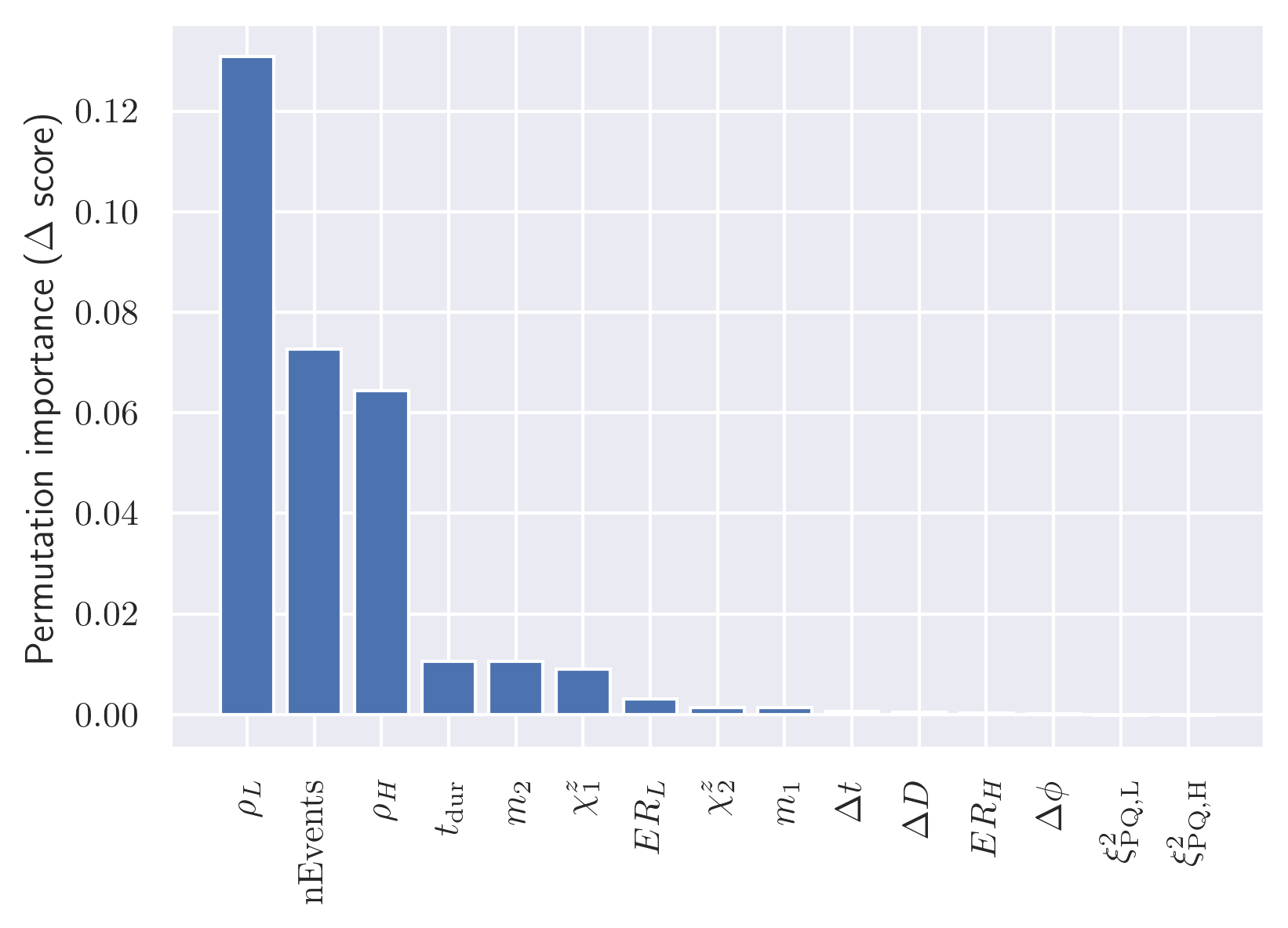}
\caption{Feature importance for the O3a model obtained via feature permutation importance.}
\label{fig:feature-importance}
\end{figure}

Also, we explore the impact of the features in the classifier. The idea is to check which feature is most relevant for the algorithm to classifying the data --- assuming a well-performing forest, as in this case --- with an inspection technique that measures the contribution of each feature to a fitted model’s statistical performance on a given tabular dataset. We choose the feature importance permutation method. The idea behind this procedure is to randomly shuffling the values of a single feature and observing the resulting degradation of the model’s score ($\Delta$ score). By breaking the relationship between the feature and the target, we determine how much the model relies on such particular feature \cite{breiman1}. For this purpose, we used the implemented codes in \texttt{scikit-learn}. Results are reported in Fig. \ref{fig:feature-importance}.

 The largest contribution arise from the $\rho$ features (signal-to-noise ratio), which directly encode the loudness of the trigger and therefore strongly affect the classifier output. The number of events (nEvents) also shows a significant importance, reflecting the background activity of the interferometers and providing a key discriminator between noise and injection-like triggers. Conversely, the physical template parameters (component masses, spins and template duration) play a subdominant role in classification, while the statistical features exhibit a negligible impact on predictive performance, white the notable exception of $L_\mathrm{ER}$. This behavior indicates that the model primarily relies on signal loudness and background-related information rather than detailed source parameters.

\section{Application of the New Statistic: $p_\mathrm{astro}$}

The statistics $p_s$ obtained from the classifier can be used to assess the probability that a given trigger is of astrophysical origin \cite{pAstro1}, \cite{pAstro2Farr}. In particular this can be defined as the ratio between the rate density of signals $R(k)_S$ and the rate of all the triggers detected, that corresponds to the sum  $R(k)_S + R(k)_N$, given the latter term as the rate density of noise triggers, where $k$ is the ranking-statistics. Formally $p_\mathrm{astro}$ is

\begin{equation}
	p_\mathrm{astro} = \frac{R(k)_S}{R(k)_S + R(k)_N}
\end{equation}

In general, these rate densities vary as a function of $k$, which in this case is represented by the classifier score $p_s$ produced by the algorithm. The signal and noise rate densities can be written as:

\begin{align}
	R(k)_S = & p(p_s | s) \Lambda_1 \\
	R(k)_N = & p(p_s | n) \Lambda_0 
\end{align}

These rate densities have been factorized considering the effective rate, that is the expected number of astrophysical events during the data acquisition campaign's time, $\Lambda_{1}$ (priors)  or the rate of noise triggers $\Lambda_0$ \cite{ratesAndPops}, each one multiplied for the probability density functions, $p(p_s | s)$ for signals and  $p(p_s | n)$ for noise. Those functions, once knew, will quantify the distribution of 'Noise' and 'Injections' triggers among the statistics $p_s$.

The $p_\mathrm{astro}$ results finally in this expression

\begin{equation}
	p_\mathrm{astro} = \frac{ p(p_s | s) \Lambda_1}{ p(p_s | s) \Lambda_1 +  p(p_s | n) \Lambda_0}
\end{equation}

The probability density functions (PDFs) used in the $p_{\text{astro}}$ calculation can be constructed from the distribution of $p_s$ values assigned to the test dataset by the classifier. These PDFs are estimated using the Kernel Density Estimation (KDE) technique \cite{kde} \footnote{The KDE is a non-parametric technique that estimates a probability density function by centering a kernel --- here, a Gaussian --- at each data point and summing the results. The bandwidth controls the width of the kernel, thereby determining the smoothness of the resulting density. In this work, the bandwidth was chosen empirically to ensure a smooth and faithful reconstruction of the underlying distribution.}.
Since the distributions of noise and injection triggers in $p_s$ spans the interval $[0,1]$, a transformation is applied to the statistics to avoid boundary effects in the KDE procedure. Specifically, instead of working directly with $p_s$, we define the transformed statistic:

\begin{equation}
	\tilde{p}_s = ln \Big( \frac{p_s}{1 - p_s} \Big )
\end{equation}

This transformation (a logit function) stretches the distribution, particularly at the tails, and enhances the separation between signal and noise populations. For KDE, a Gaussian kernel is used, with empirically chosen bandwidths:

\begin{itemize}
	\item $b_s = 0.8$ for the signal population
	\item $b_n = 0.6$ for the noise population
\end{itemize}

The resulting PDFs are shown in Fig. 5. As expected, the populations show greater separation at the extreme values of $\tilde{p}_s$, while remaining overlapped around $\tilde{p}_s \approx 0$.
Triggers falling on the right side of the plot (large $\tilde{p}_s$) will be assigned $p_\mathrm{astro} \sim 1$, and thus classified as likely injections. Conversely, triggers on the left side will be labeled as noise, with $p_\mathrm{astro} \sim 0$. The blue and red curves in the figure represent the PDFs for noise and signal, respectively, as obtained via KDE.

\begin{figure}[t]
\centering
\includegraphics[width=\linewidth]{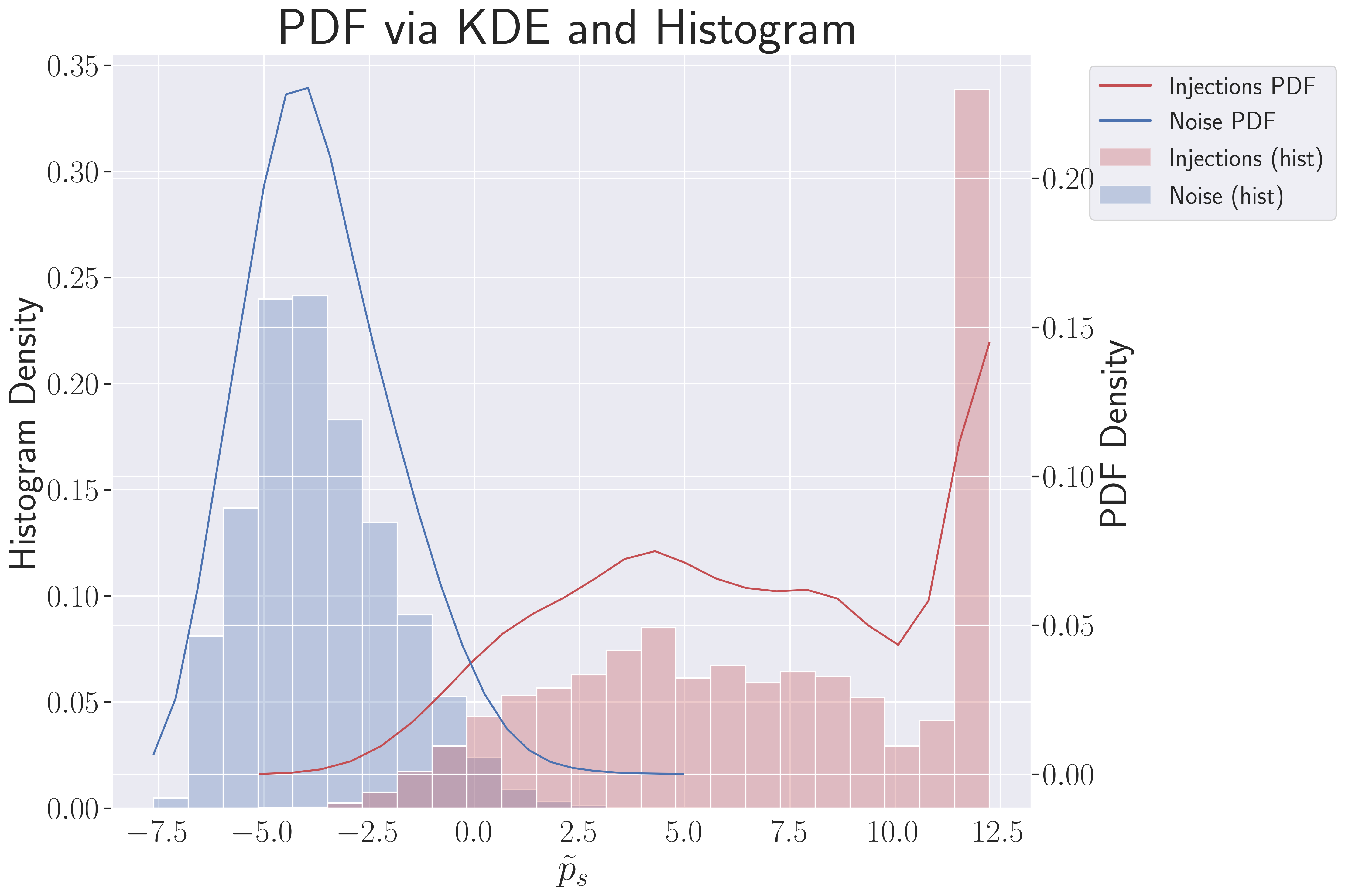}
\caption{Transformed classifier score $\tilde{p}_s$ for Noise (blue) and Injections (red) in the O3a test dataset.}
\label{fig:pdf_final}
\end{figure}

\begin{figure}[t]
\centering
\includegraphics[width=\linewidth]{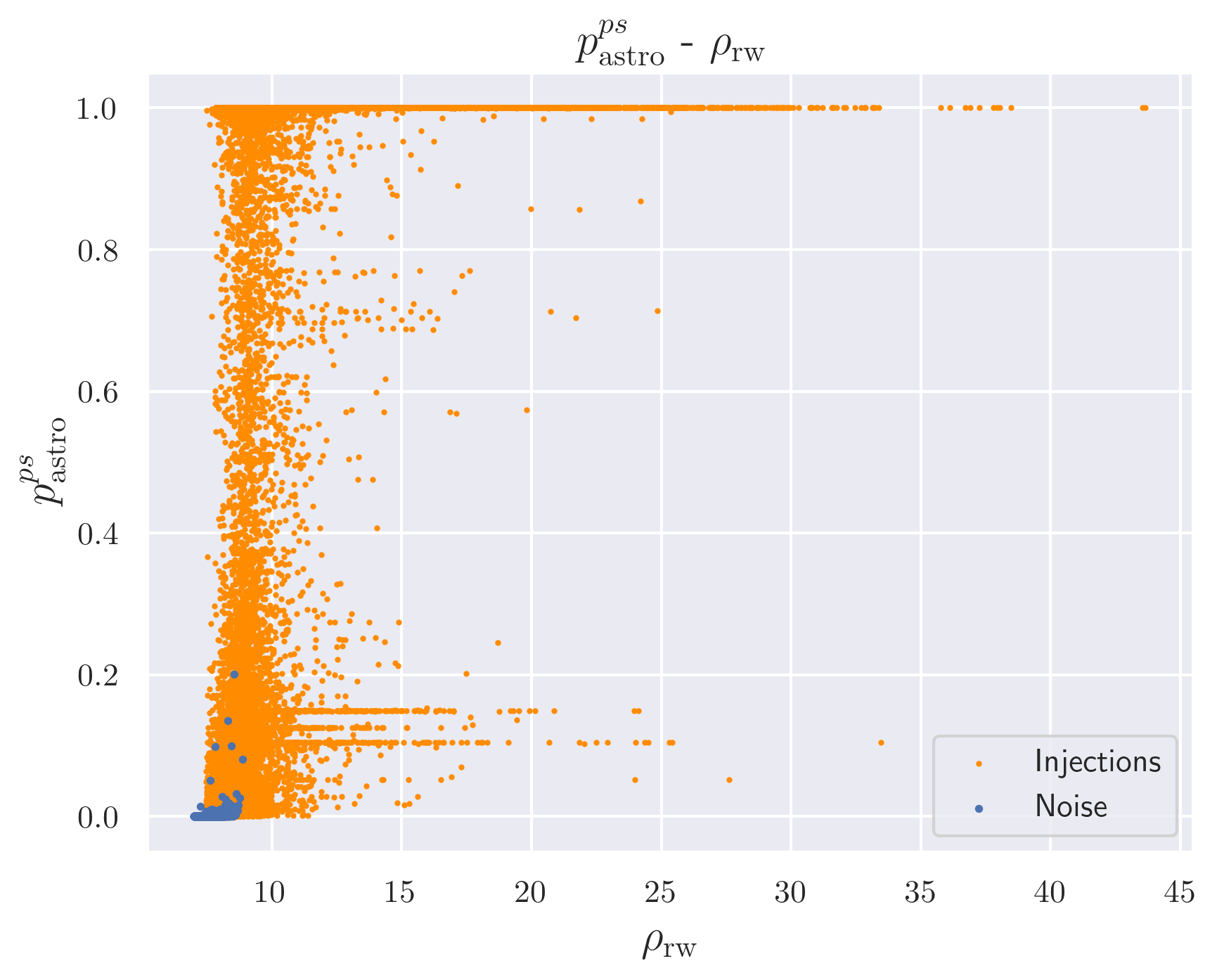}
\caption{Distribution of $p_\mathrm{astro}^{(p_s)}$ for Noise (red) and Injections (blue) in the O3a dataset.}
\label{fig:pAstro_ps}
\end{figure}

	With the estimated PDFs, it is now possible to compute $p_{\mathrm{astro}}$. We define the probability obtained using the transformed classifier statistic $\tilde{p}s$ as $p_\mathrm{astro}^{(p_s)}$. The values chosen for the signal priors are $\Lambda_{1,\mathrm{O3a}} = 39$ for O3a and $\Lambda_{1,\mathrm{O3b}} = 35$ for O3b. These values correspond to the number of triggers with an Inverse False Alarm Rate (IFAR) $> 0.5$ years, as reported by the LIGO–Virgo–KAGRA Collaboration during the O3a data acquisition campaigns \cite{gwtc2}, while for O3b this number correspond to the triggers with $p_\mathrm{astro} > 0.5$ \cite{gwtc3}.
	The choice of these prior values is conservative, as it considers only the significant events reported during the observing runs and ignores all possible subthreshold signals—i.e., real but marginally significant events that do not meet the catalogue inclusion threshold. An alternative approach, suggested in \cite{pAstro2Farr},\cite{supplement}, is to estimate the signal rate prior via self-consistent inference, using a likelihood constructed from the observed trigger distribution.
In Fig. \ref{fig:pAstro_ps}, we show the values of $p_{\mathrm{astro}}^{(p_s)}$ computed from $\tilde{p}s$ for both noise and injection triggers in the test dataset, plotted as a function of the MBTA ranking statistic, which is the quantity used by MBTA to estimate $p_\mathrm{astro}$.

	All noise triggers show $p_\mathrm{astro}^{(p_s)} < 0.15$, while the injection triggers span the full range of $p_\mathrm{astro}^{(p_s)}$ values, from 0 to 1. An encouraging observation is the correlation between $p_\mathrm{astro}^{(p_s)}$ and the MBTA ranking-statistic, suggesting that the classifier produces results consistent with those of MBTA when evaluating the astrophysical significance of triggers.
	
	However, it is important to note that several injection triggers with high ranking-statistics exhibit unexpectedly low values of $p_\mathrm{astro}^{(p_s)}$. This behavior is puzzling, especially considering that all noise triggers have amplitudes below 10, and the classifier is trained using features that fully capture the information used in the ranking-statistics calculation. Further investigation is required to understand the origin of this discrepancy. 
	Additionally, the presence of 'horizontal lines' at specific values of $p_\mathrm{astro}^{(p_s)}$ arises from the KDE procedure. Specifically, the bandwidth of the kernel smooths the distribution at certain rank values, causing several $p_s$ to map the same $p_\mathrm{astro}^{(p_s)}$ values. This effect is evident in Fig. \ref{fig:pAstro_ps}, where the lack of a  one-to-one ranking-statistics $p_s$ becomes apparent. In principle, such a one-to-one mapping is not expected. 

	To compare our results with a metric already used for detection claims, we considered the number of recovered injection triggers with $p_\mathrm{astro}^{(p_s)} > 0.5$ and compared it to the number of injections with $IFAR > 0.5$ years, which is the threshold commonly adopted to assess whether an injection can be regarded a confident detection.
	For the O3a dataset, we find that the $13,682$ out of the $24,997$ injections in the test set exceed the $p_\mathrm{astro}^{(p_s)}$ threshold, while $13,814$ injections surpass the $IFAR$ threshold. In contrast, for O3b, $13,740$ injection triggers are found above the $p_\mathrm{astro}^{(p_s)}$ threshold, whereas $12,635$ exceed $IFAR$ threshold, out of a total of $22,917$ injections in test dataset. 

\section{Analysis of Catalogued Events}

\begin{figure}
\centering
\includegraphics[width=\linewidth]{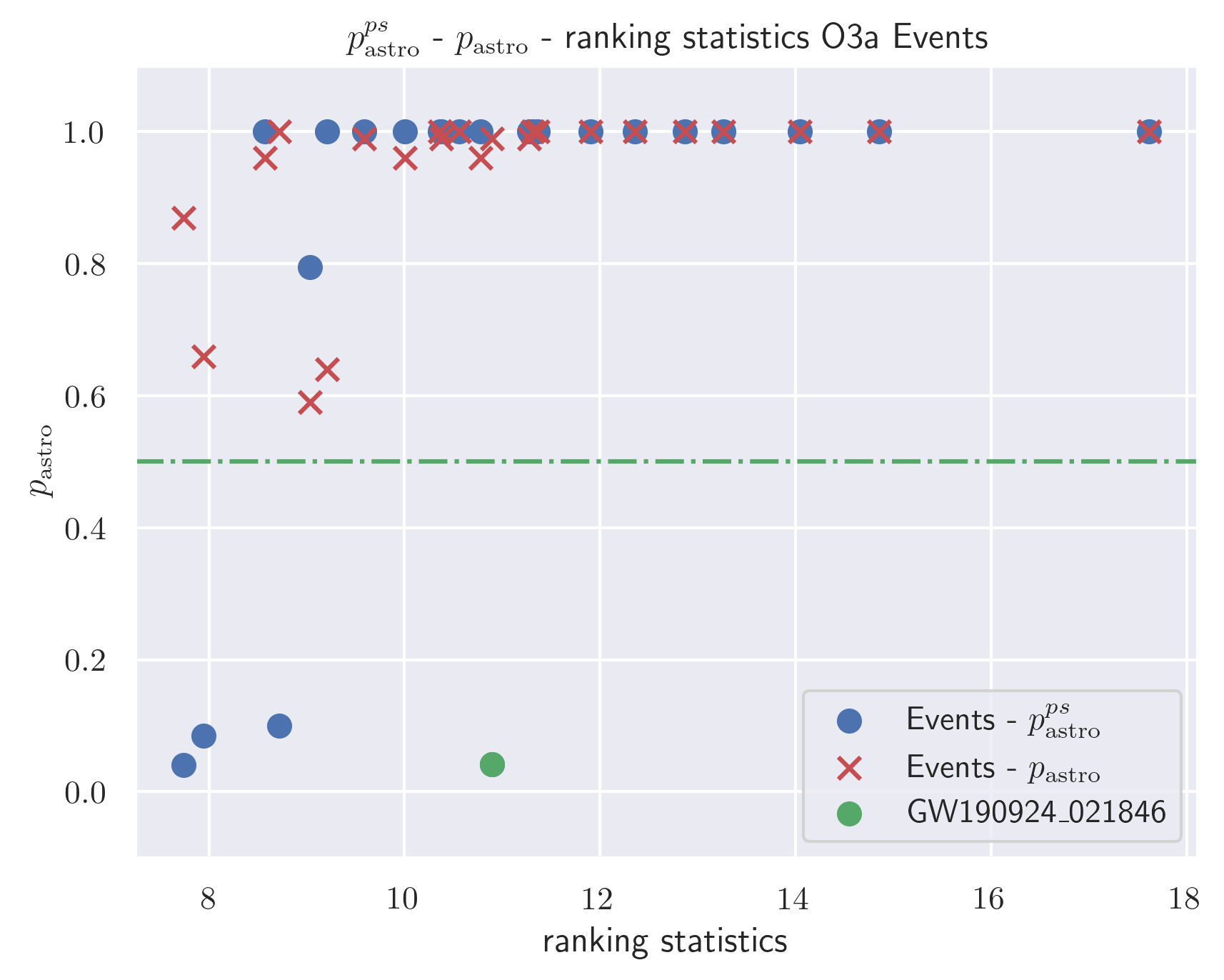}
\includegraphics[width=\linewidth]{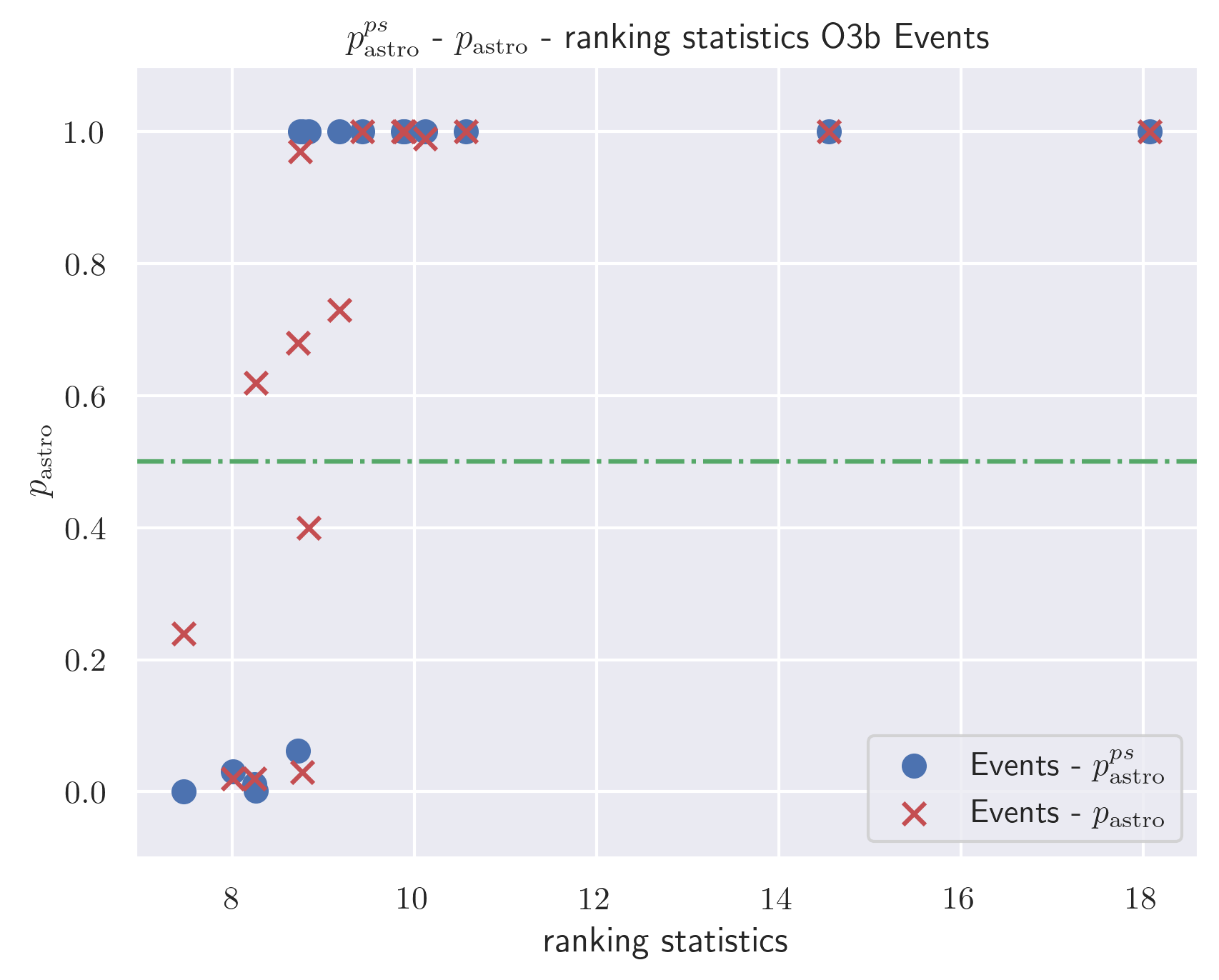}
\caption{Comparison of $p_\mathrm{astro}$ values derived from MBTA and Random Forest statistics: 
(a) O3a events from the GWTC-2.1 catalog; 
(b) O3b events from the GWTC-3 catalog.}
\label{fig:pAstro_ps_O3a_O3b}
\end{figure}

	As reported in \cite{gwtc2} and \cite{gwtc3}, the MBTA pipeline identified 23 coincidence events in O3a, and 16 events in O3b. Among these, in O3b, there are five events with $p_\mathrm{astro} < 0.5$ that were nonetheless included in the catalogues. These five events were retained in the current analysis to investigate whether the Random Forest classifier is capable of assigning them higher $p_\mathrm{astro}^{(p_s)}$ values.
	To perform the comparison, the trained models and the derived PDFs were applied to the catalogued events, and the corresponding $p_\mathrm{astro}^{(p_s)}$ values were computed. As shown in Fig. \ref{fig:pAstro_ps_O3a_O3b}, the majority of events have $p_\mathrm{astro}^{(p_s)}$ values consistent with those obtained from the MBTA analysis—particularly for events with ranking statistic values above $10$.
	In O3a, for ranking statistics below $10$, there are three events with $p_\mathrm{astro}^{(p_s)} < 0.5$, and only one of them has a ranking statistics greater than $10$ --- namely, the GW190924$\_$021846 event. This case will be discussed in detail in the next subsection.
	In O3b, for ranking statistics below $10$, there are five events with $p_\mathrm{astro}^{(p_s)} < 0.5$ with no anomalous behavior for events with ranking statistics above $10$. \footnote{We stress that some triggers show  $p_\mathrm{astro}$ values below thresholds in other pipelines apart from MBTA. In particular: GW190916\_200658 has $p_\mathrm{astro} = 0.08$ (GstLAL), GW200208\_2222617 shows $p_\mathrm{astro} = 0.01$ (GstLAL) and GW200322\_091133 $p_\mathrm{astro} = 0.08$ (PyCBC).} 

\subsection{Case Study: GW190924\_021846}

\begin{figure}[t]
\centering
\includegraphics[width=\linewidth]{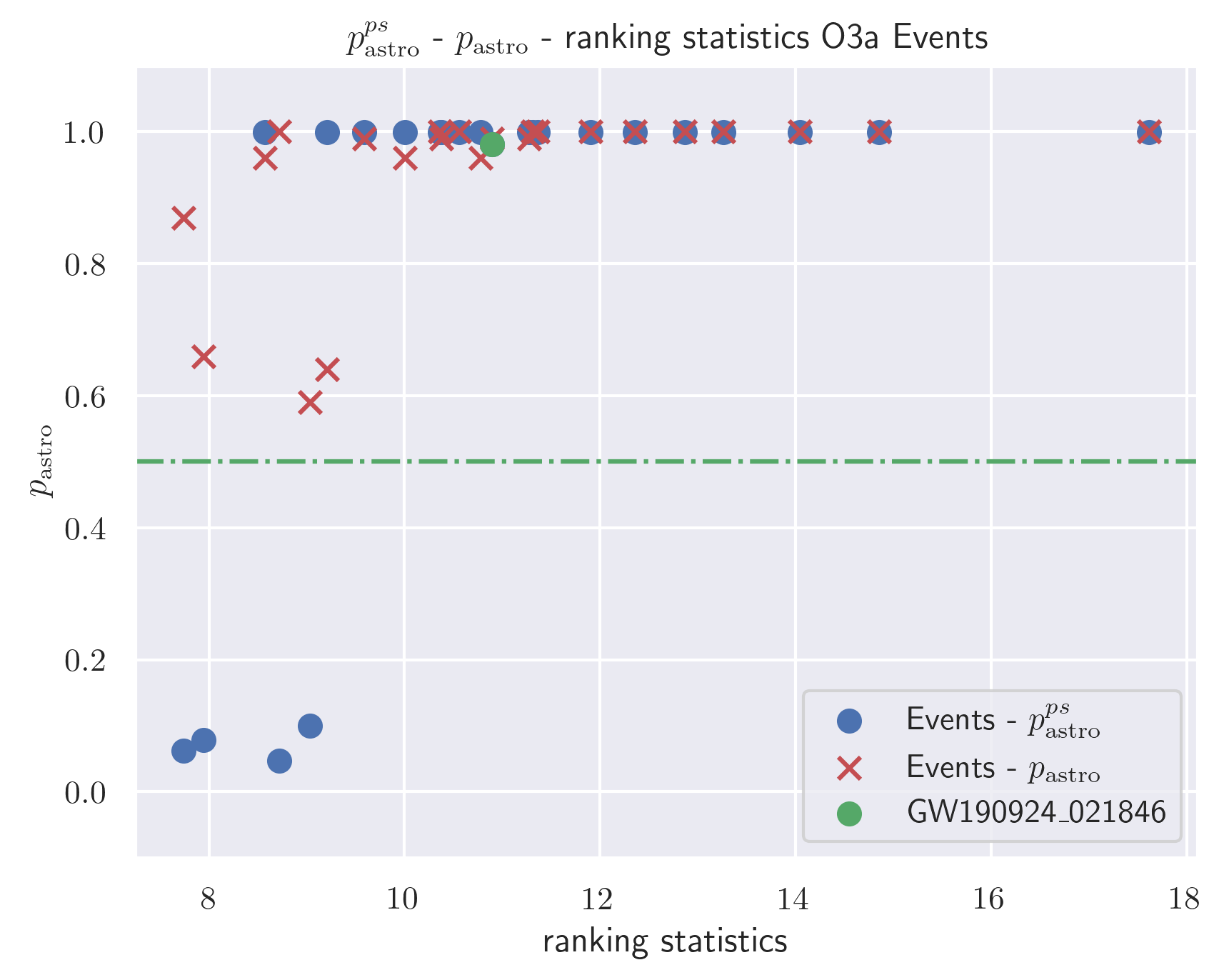}
\caption{Distribution of $p_\mathrm{astro}$ for O3a GWTC-2.1 catalog events, excluding $ER$ features.}
\label{fig:pAstro_ps_O3a_ER}
\end{figure}

The event GW190924$\_$021846 is classified as a vanilla binary black hole (BBH) merger, with source-frame component masses reported by GWOSC \cite{gwosc} as $m_1 = 80.8^{+33.0}_{-33.2} , M\odot$ and $m_2 = 24.1^{+19.3}_{-10.6} , M\odot$. The corresponding MBTA trigger does not exhibit any apparent anomalies, and the online parameters provided by the pipeline are $m_1 = 41.3$, $m_2 = 1.97$, with a ranking statistic value of 10.9. However, the classifier unexpectedly fails to assign a high score to this event. Specifically, the value of $p_\mathrm{astro}^{(p_s)}$ is only 0.04, which is inconsistent with the event's amplitude and significance. Upon further inspection, it was empirically observed that removing the $ER_H$ and $ER_L$ features from the input feature set led to a substantial improvement in the classification of this event. As shown in Fig. \ref{fig:pAstro_ps_O3a_ER}, the $p_\mathrm{astro}^{(p_s)}$ score shifts from 0.04 to $p_\mathrm{astro}^{(p_s, noER)} = 0.98$. This dramatic change suggests that the excess rate features ($ER_H$, $ER_L$) heavily bias the classifier against this particular event. It is important to highlight that in doing so, event GW190725$\_$174728 falls from $0.77$ to $0.09$, even though this event has a ranking-statistics below $10$, causing this fall more comprehensible despite this loss in significance. The reasons for this behavior are not yet understood and requires further investigations, especially given that the event GW190924$\_$021846 otherwise exhibits high significance and physical plausibility.

For completeness, in Tables \ref{table:pastro_O3a} and \ref{table:pastro_O3b} all the, $p_\mathrm{astro}$ and $p_\mathrm{astro}^{(p_s,\mathrm{noER})}$, that are the $p_\mathrm{astro}^{(p_s)}$ values calculated without the $ER$ features, are reported for all the GWTC-2.1 and GWTC-3 HL-Von MBTA events.

\begin{table}[t]
\caption{O3a events $p_\mathrm{astro}$ comparison.}
\label{table:pastro_O3a}
\centering
\begin{ruledtabular}
\begin{tabular}{lcccc}
Event name & ranking-statistics & $p_\mathrm{astro}$ & $p_\mathrm{astro}^{(p_s,\mathrm{ER})}$ & $p_\mathrm{astro}^{(p_s,\mathrm{noER})}$ \\
\colrule
GW190408\_181802 & 14.05 & 1.00 & 1.000 & 0.999 \\
GW190412 & 17.62 & 1.00 & 1.000 & 0.999 \\
GW190413\_134308 & 9.59 & 0.99 & 1.000 & 0.999 \\
GW190503\_185404 & 11.36 & 1.00 & 1.000 & 0.999 \\
GW190512\_180714 & 11.28 & 0.99 & 1.000 & 0.999 \\
GW190513\_205428 & 10.38 & 0.99 & 1.000 & 0.999 \\
GW190517\_055101 & 10.37 & 1.00 & 1.000 & 0.999 \\
GW190519\_153544 & 13.26 & 1.00 & 1.000 & 0.999 \\
GW190521 & 10.78 & 0.96 & 1.000 & 0.999 \\
GW190602\_175927 & 11.91 & 1.00 & 1.000 & 0.999 \\
GW190701\_203306 & 7.74 & 0.87 & 0.036 & 0.055 \\
GW190706\_222641 & 11.31 & 1.00 & 1.000 & 0.999 \\
GW190720\_000836 & 10.56 & 1.00 & 1.000 & 0.999 \\
GW190725\_174728 & 9.03 & 0.59 & 0.775 & 0.090 \\
GW190728\_064510 & 12.86 & 1.00 & 1.000 & 0.999 \\
GW190803\_022701 & 8.57 & 0.96 & 1.000 & 0.999 \\
GW190828\_063405 & 14.85 & 1.00 & 1.000 & 0.999 \\
GW190828\_065509 & 10.00 & 0.96 & 1.000 & 0.999 \\
GW190915\_235702 & 12.35 & 1.00 & 1.000 & 0.999 \\
GW190916\_200658 & 7.94 & 0.66 & 0.076 & 0.070 \\
GW190929\_012149 & 9.21 & 0.64 & 1.000 & 0.999 \\
GW190727\_060333 & 8.72 & 1.00 & 0.090 & 0.042 \\
GW190924\_021846 & 10.89 & 0.99 & 0.037 & 0.978 \\
\end{tabular}
\end{ruledtabular}
\end{table}

\begin{table}[t]
\caption{O3b events $p_\mathrm{astro}$ comparison.}
\label{table:pastro_O3b}
\centering
\begin{ruledtabular}
\begin{tabular}{lcccc}
Event name & ranking-statistics & $p_\mathrm{astro}$ & $p_\mathrm{astro}^{(p_s,\mathrm{ER})}$ & $p_\mathrm{astro}^{(p_s,\mathrm{noER})}$ \\
\colrule
GW191105\_143521 & 9.89 & 1.00 & 1.000 & 1.000 \\
GW191113\_071753 & 8.73 & 0.68 & 0.063 & 1.000 \\
GW191127\_050227 & 9.18 & 0.73 & 1.000 & 1.000 \\
GW191215\_223052 & 10.12 & 0.99 & 1.000 & 1.000 \\
GW191230\_180458 & 8.85 & 0.40 & 1.000 & 1.000 \\
GW200115\_042309 & 10.56 & 1.00 & 1.000 & 1.000 \\
GW200208\_130117 & 9.43 & 1.00 & 1.000 & 1.000 \\
GW200208\_222617 & 8.25 & 0.02 & 0.011 & 0.007 \\
GW200209\_085452 & 8.75 & 0.97 & 1.000 & 1.000 \\
GW200216\_220804 & 8.01 & 0.02 & 0.031 & 0.060 \\
GW200219\_094415 & 9.88 & 1.00 & 1.000 & 1.000 \\
GW200224\_222234 & 18.07 & 1.00 & 1.000 & 1.000 \\
GW200308\_173609 & 7.47 & 0.24 & 0.000 & 0.000 \\
GW200311\_115853 & 14.55 & 1.00 & 1.000 & 1.000 \\
GW200316\_215756 & 8.77 & 0.03 & 1.000 & 1.000 \\
GW200322\_091133 & 8.26 & 0.62 & 0.001 & 0.001 \\
\end{tabular}
\end{ruledtabular}
\end{table}

\section{Search for New Candidates}

Since this method shows consistency with the results obtained by the standard MBTA pipeline, an agnostic search for new candidate events was carried out. Both O3a and O3b datasets were fully analyzed. For this search, the complete dataset was split into two equal subsets, with careful attention paid to maintaining class balance between signal and noise triggers in each subset. For model training, $70\%$ of the training subset was used to train the classifier, while the remaining $30\%$ was used for validation, specifically to estimate the probability density functions (PDFs) following the procedure described in Sec. 5. The hyperparameters used were identical to those employed in the standard (previous) analysis.
Once the model was trained and the PDFs were constructed, it was applied to the test subset, and the $p_\mathrm{astro}^{(p_s,\mathrm{noER})}$ values were computed. If a noise trigger was found to have $p_\mathrm{astro}^{(p_s,\mathrm{noER})} > 0.5$, it was promoted to a possible candidate. This entire process was repeated with the roles of the training and test datasets inverted to ensure coverage of the full dataset. As a result of this search, one candidate was identified with $p_\mathrm{astro}^{(p_s,\mathrm{noER})} = 0.92$, at GPS time 1240423628.7, with a ranking statistic $\rho_{\mathrm{rw,ER}} = 8.88011$ and $\text{IFAR} = 0.05$ years. The MBTA trigger is associated to a BBH system of $m_1 = 24.4 M_\odot$ and $m_2 = 4.8 M_\odot$. This event is also reported in \cite{3-OGC} with $p_\mathrm{astro} = 0.41$ as GW190427$\_$180650, while it is not present in the official catalogue as low-significance trigger. 

\section{Conclusions}

This work investigated the application of machine learning techniques to gravitational-wave data analysis. By considering a diverse set of features, a Random Forest classifier was developed and extensively optimized through the exploration of a broad range of hyperparameter combinations. Once trained, the classifier was evaluated, and the resulting statistic ($p_s$) was compared to the ranking statistic currently used by the MBTA pipeline. In addition, we examined the relative importance of the input features for this classification task, finding that some features contribute only marginally to the model’s decision process, suggesting potential redundancy rather than clear necessity for accurate trigger classification.

The results demonstrated a consistent result in detection efficiency when using the Random Forest–based approach. A further application was explored by employing the $p_s$ statistic to construct a $p_\mathrm{astro}$ probability distribution, which quantifies the likelihood of a trigger being astrophysical in origin. To achieve this, probability density functions were estimated using the KDE technique, applied to the test dataset after applying a logit transformation to the $p_s$ values to enhance the KDE performance.
Using this derived metric, we compared the number of injection triggers that exceeding the threshold $p_\mathrm{astro}^{(p_s)} > 0.5$ and $IFAR > 0.5$ years, obtaining comparable results for O3a and slightly improved performance for O3b. These outcomes are encouraging since the $IFAR$ computation relies on background modelling, which can be time and resource computing, whereas the $p_\mathrm{astro}^{(p_s)}$ metric is obtained in a more straightforward manner. Furthermore, the $p_\mathrm{astro}^{(p_s)}$ values were derived and compared to those obtained from the MBTA pipeline ($p_\mathrm{astro})$, considering all events reported in the O3a and O3b catalogs. 
The results were generally consistent, with the notable exception of GW190924$\_$021846, which exhibited unexpected behavior due to the influence of specific features in the model. This prompted a deeper analysis of the feature–tree interactions within the classifier. Finally, an agnostic search for new candidate events was conducted across the full O3a and O3b datasets. This search yielded one new candidate, with $p_\mathrm{astro}^{(p_s,\mathrm{noER})} = 0.97$,
marking a promising outcome for the use of machine learning–based tools in gravitational-wave event classification.

The Random Forest classifier demonstrated the ability to construct a statistic with a level of separability between noise and signal triggers that is comparable to that achieved using the standard ranking statistic. Moreover, the application of the classifier output to compute $p_\mathrm{astro}^{(p_s)}$ is promising, suggesting that this method could serve as a reliable tool for estimating astrophysical detection probabilities.
As future work, alternative approaches may be explored, including the use of unsupervised learning techniques, such as Autoencoders \cite{autoencoder}, for denoising purposes. Another potential direction involves leveraging the classifier to directly estimate the False Alarm Rate (FAR) associated with triggers, offering further enhancements to real-time detection pipelines.

\section*{Data Availability}

The data and analysis scripts supporting this work are available in Ref.~\cite{mobilia2026rf4gw}, with the development repository hosted on GitHub.

\begin{acknowledgments}
We thank Thomas Dent (IGFAE, University of Santiago de Compostela) for a thorough and helpful review and Tito Dal Canton (IJC Lab - Université Paris-Saclay) for his valuable suggestions. We also thank the MBTA team for their continued advice and support, especially Florian Aubin and Thomas Sainrat (Université de Strasbourg) for the final review of this document. This research has made use of data or software obtained from the Gravitational Wave Open Science Center (gwosc.org), a service of the LIGO Scientific Collaboration, the Virgo Collaboration, and KAGRA. This material is based upon work supported by NSF’s LIGO Laboratory which is a major facility fully funded by the National Science Foundation, as well as the Science and Technology Facilities Council (STFC) of the United Kingdom, the Max-Planck-Society (MPS), and the State of Niedersachsen/Germany for support of the construction of Advanced LIGO and construction and operation of the GEO600 detector. Additional support for Advanced LIGO was provided by the Australian Research Council. Virgo is funded, through the European Gravitational Observatory (EGO), by the French Centre National de Recherche Scientifique (CNRS), the Italian Istituto Nazionale di Fisica Nucleare (INFN) and the Dutch Nikhef, with contributions by institutions from Belgium, Germany, Greece, Hungary, Ireland, Japan, Monaco, Poland, Portugal, Spain. KAGRA is supported by Ministry of Education, Culture, Sports, Science and Technology (MEXT), Japan Society for the Promotion of Science (JSPS) in Japan; National Research Foundation (NRF) and Ministry of Science and ICT (MSIT) in Korea; Academia Sinica (AS) and National Science and Technology Council (NSTC) in Taiwan.
\end{acknowledgments}

\appendix
\section{Feature distributions}
\begin{figure*}[t]
    \centering
    \includegraphics[width=\textwidth]{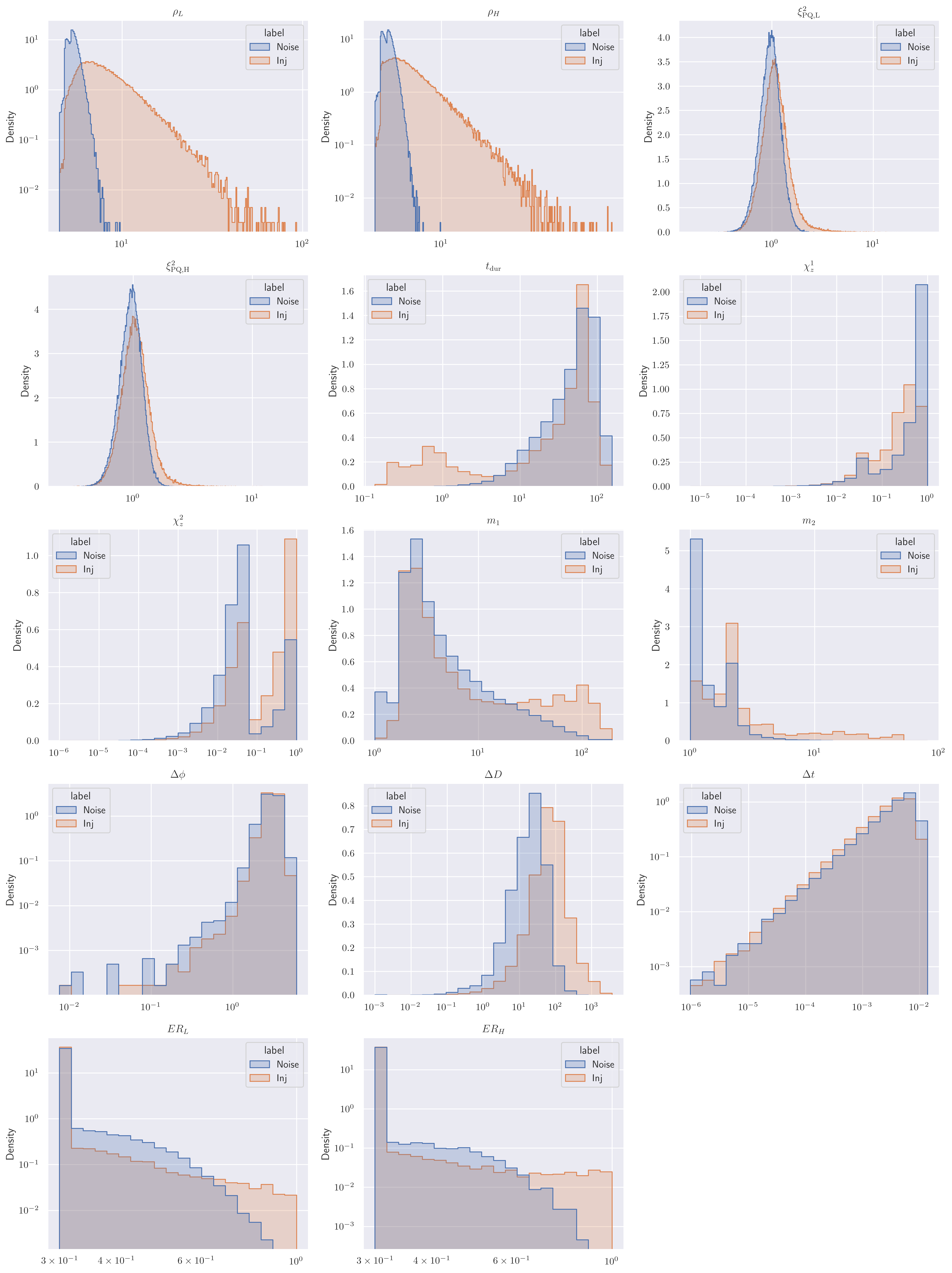}
    \caption{Distribution of the features for noise (blue) and injections (red) populations.}
    \label{fig:feature_distributions}
\end{figure*}

\begin{figure*}[t]
    \centering
    \includegraphics[width=\textwidth]{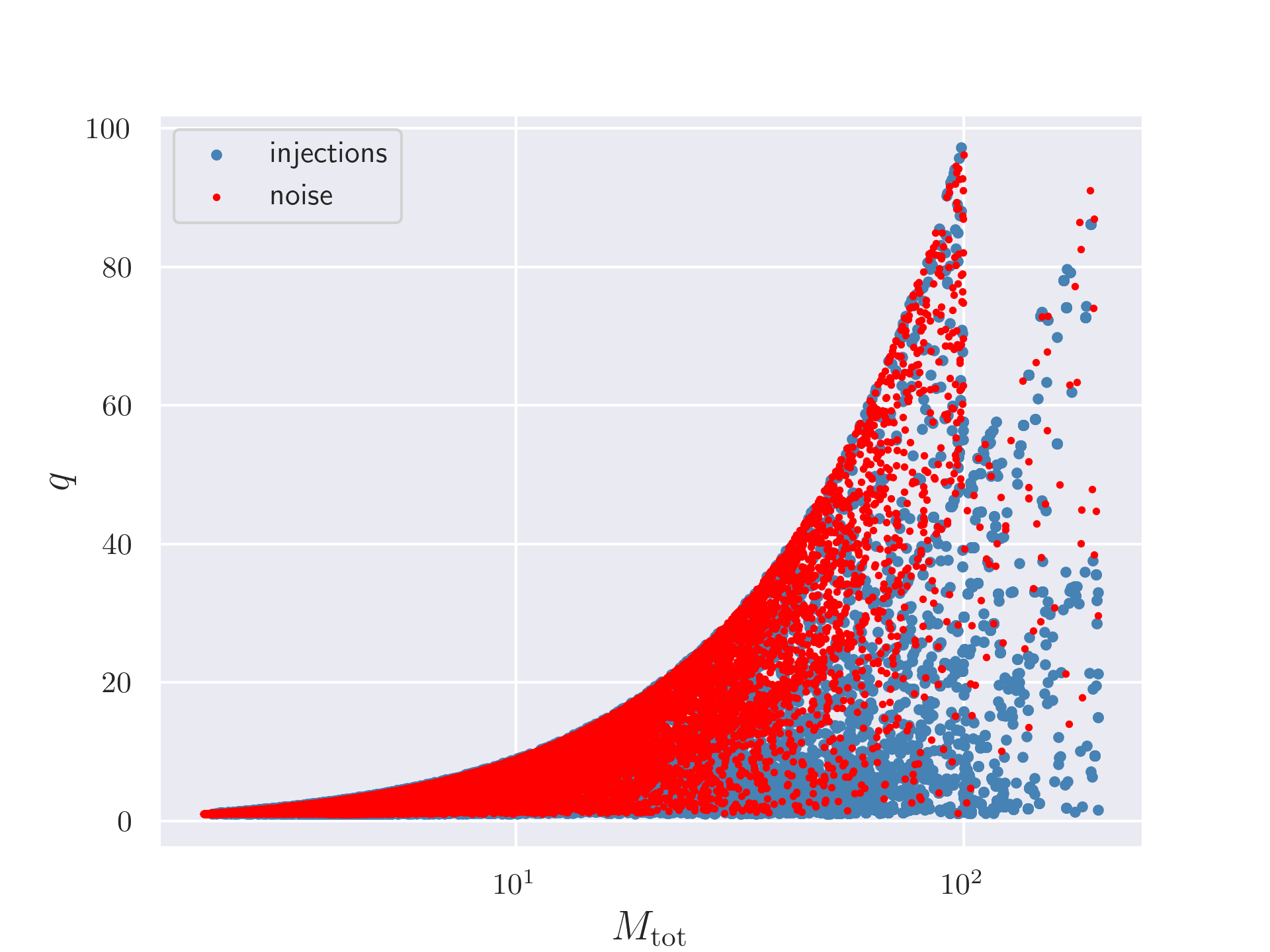}
    \caption{Distribution of masses population for injections and noise labeled triggers in total mass ($M_\mathrm{tot}$) and mass-ratio ($q = \frac{m_1}{m_2}$).}
    \label{fig:mass-population}
\end{figure*}

Fig. \ref{fig:feature_distributions} shows the distribution of features described in Sec. IV and used in this study for both the injection and noise populations. Each subfigure presents the distribution for a single feature, with the feature name indicated at the top of the plot. The Random Forest algorithm leverages non-trivial correlations among these features to construct the ranking statistic $p_s$. By incorporating both physical and statistical informations, this rank is expected to outperform --- or being comparable to --- the Ranking statistic provided by MBTA. 
A focus on the mass population is also depicted in Fig. \ref{fig:mass-population}.

\label{appendix:features}

\bibliographystyle{apsrev4-2}
\bibliography{references}

\end{document}